\documentclass[
  prd,preprint,longbibliography,
  showpacs,showkeys,lengthcheck,
  nofootinbib,tightenlines,onecolumn,notitlepage,
  preprintnumbers,superscriptaddress
]{revtex4-1}

\usepackage[utf8]{inputenc}
\usepackage{newtxtext,newtxmath}
\usepackage[mathcal]{euscript}

\usepackage{graphicx}
\usepackage[usenames,dvipsnames]{xcolor}
\graphicspath{{figures/}}
\usepackage{tikz-feynman}

\usepackage{hyperref}
\hypersetup{colorlinks=true,citecolor=blue,linkcolor=blue,urlcolor=blue}
\usepackage{orcidlink}

\usepackage{amsmath}
\usepackage{slashed}
\usepackage{bm}
\usepackage{bbm}
\usepackage{mathtools}
\usepackage{tensor}

\usepackage[math]{cellspace}
\usepackage[final]{changes} 

\renewcommand*{\d}{\mathop{}\!\mathrm{d}}
\newcommand*{\e}{\mathop{}\!\mathrm{e}}

\newcommand{\change}{\delta_{\xi}}

\newcommand{\Lag}{\ensuremath{\mathscr{L}}}

\newcommand{\D}{\mathcal{D}}
\newcommand{\xp}{\tilde{x}}

\begin{document}

\title{
  Reflections on Noether's second theorem and the energy-momentum tensor
}

\author{Adam Freese \orcidlink{0000-0002-0688-4121}\,}
\email{afreese@jlab.org}
\affiliation{Center for Nuclear Femtography, Southeastern Universities Research Association, Newport News, Virginia 23606, USA}
\affiliation{Theory Center, Jefferson Lab, Newport News, Virginia 23606, USA}

\begin{abstract}
  Through symmetry of the action under global spacetime translations,
  Noether's first theorem infamously entails an energy-momentum tensor (EMT)
  that is neither symmetric nor gauge-invariant.
  In a prior work~\cite{Freese:2021jqs},
  I had obtained a symmetric and gauge-invariant EMT by using
  Noether's second theorem instead,
  with local spacetime translations as the symmetry group.
  However, the derivation therein was flawed, containing a faulty
  assumption about the transformation rule for spinor fields.
  In this work, I revisit the derivation of Ref.~\cite{Freese:2021jqs},
  both correcting the faulty step
  and simplifying the derivation for broader accessibility.
  The end result is an EMT for quantum chromodynamics
  that is gauge-invariant, but not symmetric.
\end{abstract}

\preprint{JLAB-THY-25-4355}

\maketitle


\section{Introduction}
\label{sec:intro}

The energy-momentum tensor (EMT) has become a hot topic in hadron physics,
promising to elucidate longstanding questions about
dynamical mass generation in quantum
chromodynamics~\cite{Ji:1995sv,Lorce:2017xzd,Metz:2020vxd,Ji:2021mtz,Lorce:2021xku}
and
the breakdown of the proton's spin~\cite{Jaffe:1989jz,Ji:1996ek,Leader:2013jra},
and possibly even to provide spatial distributions
of stresses experienced by
quarks and gluons~\cite{Polyakov:2002yz,Polyakov:2018zvc,Lorce:2018egm}\footnote{
  The concept of stresses in hadrons has been questioned in
  Refs.~\cite{Ji:2021mfb,Ji:2022exr,Ji:2025gsq},
  and in turn defended in
  Refs.~\cite{Burkert:2023wzr,Freese:2024rkr,Lorce:2025oot}.
}.
A desire to obtain the mechanical form factors appearing in
matrix elements of the EMT has motivated experimental studies
of
deeply virtual Compton scattering~\cite{Burkert:2018bqq,Kumericki:2019ddg,Burkert:2021ith}
and
hard exclusive meson production~\cite{Duran:2022xag},
feasibility studies for future measurements of near-threshold
meson production at Jefferson Lab and the Electron Ion Collider~\cite{Guo:2025jiz,Hatta:2025vhs},
and lattice QCD computations of these form
factors~\cite{Shanahan:2018pib,Shanahan:2018nnv,Pefkou:2021fni,Hackett:2023rif,Pefkou:2023okb,Hackett:2023nkr}.

As important as the EMT is, there are---as of
this writing---actually two
energy-momentum tensors for quantum chromodynamics
that are in common use.
The first of these is asymmetric
under exchange of its indices:
\begin{align}
  T^{\mu\nu}
  =
  \sum_q
  \frac{i}{2}
  \bar{q}
  \gamma^{\mu}
  \overleftrightarrow{\D}^{\nu}
  q
  +
  F_a^{\mu\rho} F_{\rho}^{a\,\nu}
  -
  \eta^{\mu\nu} \Lag
  \,,
\end{align}
and the other is its symmetrization
$
\frac{1}{2}
\big(
  T^{\mu\nu}
  +
  T^{\nu\mu}
\big)
$.
The latter is usually called the Belinfante EMT~\cite{Belinfante:1939emt}.

Whether the EMT is symmetric or asymmetric has implications
for properties of hadrons.
Most significantly,
the symmetric EMT justifies breaking a hadron's spin down into
quark and gluon contributions only,
while the asymmetric EMT allows the quark contribution to be further broken down
into orbital and intrinsic spin
parts ~\cite{Leader:2013jra,Lorce:2025pxt,Won:2025dgc}.
The antisymmetric part of the EMT also contributes to
the light front energy density~\cite{Lorce:2018egm,Freese:2023abr},
since this is given by the off-diagonal component $T^{+-}(x)$ rather than $T^{00}(x)$.
Moreover, even in instant form coordinates,
the momentum density $T^{0i}(x)$ and the energy flux density $T^{i0}(x)$
are no longer identical for systems with fermion constituents,
including elementary fermions themselves.
This has implications for interpretations of quantum mechanics
that postulate exact particle positions as hidden variables,
as explored for non-relativistic fermions in Ref.~\cite{Freese:2025tqd}.
As such, it is pertinent to understand whether
the symmetric or asymmetric EMT gives a more fundamental description of hadron structure.

The standard procedure to derive the EMT
is outlined meticulously in Ref.~\cite{Leader:2013jra}.
One first obtains the canonical EMT
(which is not in common use)
using Noether's first theorem,
with global spacetime translations as the relevant symmetry group.
The canonical EMT is considered unphysical because it is not gauge-invariant,
so it is modified by adding a trivially conserved quantity
(the divergence of a superpotential)
to restore gauge invariance.
Doing this can produce either the symmetric or the asymmetric EMT---which
are both gauge-invariant---depending on how the superpotential is chosen.
The choice is more often made to obtain the symmetric EMT,
but this choice is arbitrary.

In hopes of avoiding such ad hoc choices,
I proposed in Ref.~\cite{Freese:2021jqs}
to use Noether's second theorem,
with local translations as the symmetry group,
to derive the EMT.
In this way, I directly obtained the symmetric Belinfante EMT.
However, the derivation therein was flawed, because
the transformation rule I used for spinor fields was erroneous.
In fact, correcting this mistake leads to exactly the opposite result:
the EMT obtained through local translation symmetry
is the \emph{asymmetric} EMT.

It is important to correct mistakes when they appear in the scientific literature,
even if these are sometimes self-corrections.
This is especially true when the correction changes the conclusion.
This paper's primary purpose is to correct the error in
Ref.~\cite{Freese:2021jqs}.
While doing so,
I also aim to improve upon the presentation of the original work,
and present a self-contained derivation of the corrected result.

This work is organized as follows.
Section~\ref{sec:theorem}
gives a lightning-quick sketch of Noether's second theorem.
Section~\ref{sec:trans}
then gives a more detailed explanation of the local translations
being considered as a symmetry group.
It is in this section I explain and correct
the flaw in Ref.~\cite{Freese:2021jqs}.
Section~\ref{sec:emt} uses local translations and Noether's second theorem
to obtain the EMT in both QED and QCD,
which both turn out to be asymmetric.
Section~\ref{sec:end} concludes the paper,
and an appendix afterwards contains a proof that the spinor transformation rule
used in Ref.~\cite{Freese:2021jqs} is not mathematically sound.


\section{Quick sketch of Noether's second theorem}
\label{sec:theorem}

In this section, I will give a brief sketch of Noether's second theorem.
I will not discuss her more popular first theorem;
an excellent, in depth-exposition thereof can be found in
Kosyakov's textbook~\cite{Kosyakov:2007qc}.
I will also limit the discussion to a specialized case,
where the matter fields transform but the spacetime coordinates do not;
see Noether's original treatment~\cite{Noether:1918zz}
for the more general case.
The treatment herein will differ from Ref.~\cite{Freese:2021jqs},
but will be significantly simpler.

Consider a field theory with an action
\begin{align}
  S
  =
  \int \d^4 x \,
  \Lag\big[
    \Psi_a(x), \partial_\mu\Psi_a(x)
    \big]
  \,,
\end{align}
where $\Psi_a(x)$ is a collection of matter fields.
Suppose this action is invariant when the matter fields are transformed:
\begin{align}
  \Psi_a(x)
  \mapsto
  \Psi_a(x)
  +
  \change \Psi_a(x)
  \,,
\end{align}
where the transformation is in some way parametrized by a function $\xi^\mu(x)$,
which has support only in a compact region of spacetime, is smooth,
and is bounded by a small number $\epsilon$,
but is otherwise arbitrary.
Noether's second theorem concerns an identity that can be derived
under these hypotheses.

Using the chain rule, and the hypothesis that $\change S=0$:
\begin{align}
  \label{eqn:chain}
  \change S
  =
  \int \d^4 x \,
  \sum_a
  \left\{
    \frac{\partial\Lag}{\partial\Psi_a}
    \change \Psi_a
    +
    \frac{\partial\Lag}{\partial(\partial_\nu\Psi_a)}
    \change (\partial_\nu\Psi_a)
    \right\}
  =
  0
  \,.
\end{align}
To linear order in $\xi$, the integrand can be written:
\begin{align}
  \label{eqn:ABC}
  \sum_a
  \left\{
    \frac{\partial\Lag}{\partial\Psi_a}
    \change \Psi_a
    +
    \frac{\partial\Lag}{\partial(\partial_\nu\Psi_a)}
    \change (\partial_\nu\Psi_a)
    \right\}
  =
  \mathscr{A}^\mu(x) \xi_\mu(x)
  +
  \mathscr{B}^{\mu\nu}(x) \partial_\mu \xi_\nu(x)
  +
  \ldots
  \,,
\end{align}
where in principle, terms with arbitrarily high derivatives of
$\xi^\mu(x)$ might appear,
but in practice the series typically terminates at the first derivative;
in fact, this will happen for the local translations considered
in this work.
Since higher-order derivatives will not appear later,
I will drop them from consideration here.
Using integration by parts:
\begin{align}
  0
  =
  \int \d^4 x \,
  \Big\{
    \mathscr{A}^\nu(x)
    -
    \partial_\mu \mathscr{B}^{\mu\nu}(x)
    \Big\}
  \xi_\nu(x)
  \,,
\end{align}
where surface terms were dropped because $\xi^\mu(x)$ has compact support by hypothesis.
Since $\xi^\mu(x)$ is arbitrary
(aside from being smooth and having compact support),
the remainder of the integrand must identically vanish.
Therefore:
\begin{align}
  \label{eqn:theorem}
  \mathscr{A}^\nu(x)
  -
  \partial_\mu \mathscr{B}^{\mu\nu}(x)
  =
  0
  \,.
\end{align}
This is Noether's second theorem.


\section{Local translations}
\label{sec:trans}

\begin{figure}
  \includegraphics[width=\textwidth]{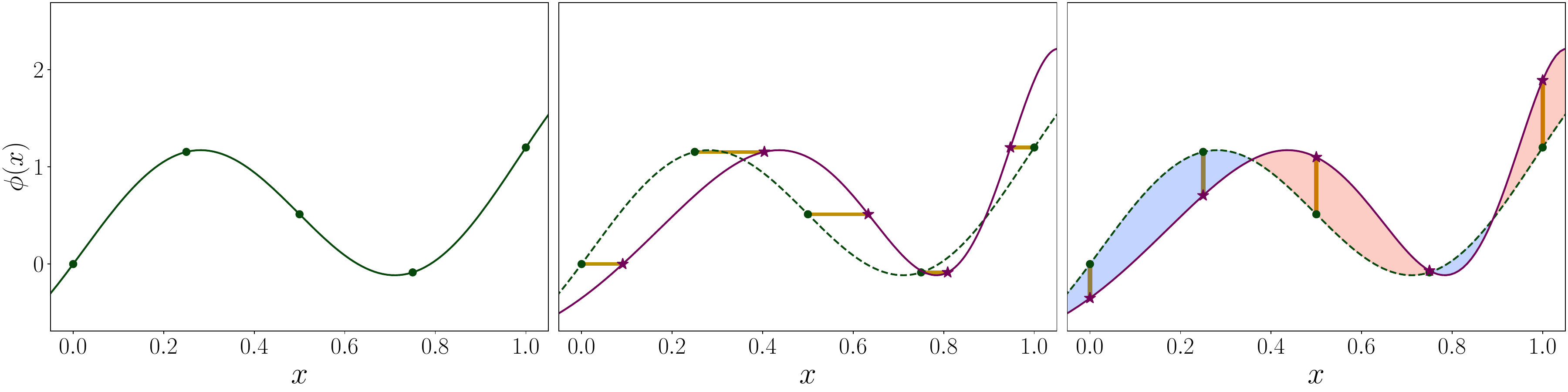}
  \caption{
    Depiction of a local translation.
    Left panel:
    a scalar function $\phi$ of one spatial variable, $x$.
    Middle panel:
    $x$ is transformed by moving every spatial point, and
    $\phi$ is reparametrized to take the same values
    at the moved points.
    Right panel:
    $\change\phi$ is evaluated by taking the difference
    between the transformed and original curve,
    per $x$ value.
  }
  \label{fig:trans}
\end{figure}

The transformation considered in this work is a local translation of the
matter fields.
What this means is that the matter fields are reparametrized
as if a general coordinate transformation had been performed,
but spacetime is not reparametrized;
see Fig.~\ref{fig:trans} for an illustration.

The tensor transformation rule for a general coordinate transformation
is~\cite{Weinberg:1972kfs,Carroll:2004st,hamilton2018general}:
\begin{align}
  \label{eqn:gct}
  {\widetilde{T}}^{\mu_1\mu_2\ldots}_{\nu_1\nu_2\ldots}(\xp)
  =
  \frac{\partial \xp^{\mu_1}}{\partial x^{\alpha_1}}
  \frac{\partial \xp^{\mu_2}}{\partial x^{\alpha_2}}
  \ldots
  \frac{\partial x^{\beta_1}}{\partial \xp^{\nu_1}}
  \frac{\partial x^{\beta_2}}{\partial \xp^{\nu_2}}
  {T}^{\alpha_1\alpha_2\ldots}_{\beta_1\beta_2\ldots}(x)
  \,,
\end{align}
where a tilde is placed over transformed quantities.
A local translation thus means the replacement:
\begin{align}
  {T}^{\mu_1\mu_2\ldots}_{\nu_1\nu_2\ldots}(x)
  \mapsto
  {\widetilde{T}}^{\mu_1\mu_2\ldots}_{\nu_1\nu_2\ldots}(x)
  \,,
\end{align}
i.e., the function is transformed, but the spacetime coordinates are not
(see Fig.~\ref{fig:trans}).
The change in the tensor field is defined:
\begin{align}
  \label{eqn:delta}
  \change
  {T}^{\mu_1\mu_2\ldots}_{\nu_1\nu_2\ldots}
  \equiv
  {\widetilde{T}}^{\mu_1\mu_2\ldots}_{\nu_1\nu_2\ldots}(x)
  -
  {T}^{\mu_1\mu_2\ldots}_{\nu_1\nu_2\ldots}(x)
  \,.
\end{align}

The next matter is finding a formula for
$
  \change
  {T}^{\mu_1\mu_2\ldots}_{\nu_1\nu_2\ldots}
$,
which requires explicit construction of the local translation.
The general coordinate transformation generating the local translation is:
\begin{align}
  \label{eqn:xi}
  x^\mu
  \mapsto
  \xp^\mu(x)
  \equiv
  x^\mu
  +
  \xi^\mu(x)
  \,,
\end{align}
where, for Noether's second theorem to be applicable,
$\xi^\mu(x)$ has compact support
and is bounded by some small parameter $\epsilon$.
For notational compactness, I will drop all order-$\xi^2$ terms,
under the rationale that they're bounded by $\epsilon^2$.
The left-hand side of Eq.~(\ref{eqn:gct}) works out to linear order as:
\begin{align*}
  {\widetilde{T}}^{\mu_1\mu_2\ldots}_{\nu_1\nu_2\ldots}(\xp)
  =
  {\widetilde{T}}^{\mu_1\mu_2\ldots}_{\nu_1\nu_2\ldots}\big(x + \xi(x)\big)
  =
  {\widetilde{T}}^{\mu_1\mu_2\ldots}_{\nu_1\nu_2\ldots}(x)
  +
  \xi^\lambda
  \partial_\lambda
  {T}^{\mu_1\mu_2\ldots}_{\nu_1\nu_2\ldots}(x)
  \,.
\end{align*}
To evaluate the right-hand side of Eq.~(\ref{eqn:gct}),
it's helpful to note that, to linear order in $\xi$:
\begin{align*}
  \begin{split}
    \frac{\partial \xp^\mu}{\partial x^\alpha}
    &=
    \delta^{\mu}_{\alpha}
    +
    \partial_\alpha \xi^\mu
    \\
    \frac{\partial x^\beta}{\partial \xp^\nu}
    &=
    \delta^{\beta}_{\nu}
    -
    \partial_\nu \xi^\beta
    \,,
  \end{split}
\end{align*}
and thus, to linear order:
\begin{multline*}
  \frac{\partial \xp^{\mu_1}}{\partial x^{\alpha_1}}
  \frac{\partial \xp^{\mu_2}}{\partial x^{\alpha_2}}
  \ldots
  \frac{\partial x^{\beta_1}}{\partial \xp^{\nu_1}}
  \frac{\partial x^{\beta_2}}{\partial \xp^{\nu_2}}
  {T}^{\alpha_1\alpha_2\ldots}_{\beta_1\beta_2\ldots}
  =
  {T}^{\mu_1\mu_2\ldots}_{\nu_1\nu_2\ldots}
  +
  (\partial_{\lambda} \xi^{\mu_1})
  {T}^{\lambda\mu_2\ldots}_{\nu_1\nu_2\ldots}
  +
  (\partial_{\lambda} \xi^{\mu_2})
  {T}^{\mu_1\lambda\ldots}_{\nu_1\nu_2\ldots}
  +
  \ldots
  \\
  -
  (\partial_{\nu_1} \xi^{\lambda})
  {T}^{\mu_1\mu_2\ldots}_{\lambda\nu_2\ldots}
  -
  (\partial_{\nu_2} \xi^{\lambda})
  {T}^{\mu_1\mu_2\ldots}_{\nu_1\lambda\ldots}
  -
  \ldots
  \,.
\end{multline*}
Putting both hands together, this tells us:
\begin{align}
  \label{eqn:delta:linear}
  \change
  {T}^{\mu_1\mu_2\ldots}_{\nu_1\nu_2\ldots}
  =
  -
  \Big\{
    \xi^\lambda
    \partial_\lambda
    {T}^{\mu_1\mu_2\ldots}_{\nu_1\nu_2\ldots}
    -
    (\partial_{\lambda} \xi^{\mu_1})
    {T}^{\lambda\mu_2\ldots}_{\nu_1\nu_2\ldots}
    -
    (\partial_{\lambda} \xi^{\mu_2})
    {T}^{\mu_1\lambda\ldots}_{\nu_1\nu_2\ldots}
    -
    \ldots
    +
    (\partial_{\nu_1} \xi^{\lambda})
    {T}^{\mu_1\mu_2\ldots}_{\lambda\nu_2\ldots}
    +
    (\partial_{\nu_2} \xi^{\lambda})
    {T}^{\mu_1\mu_2\ldots}_{\nu_1\lambda\ldots}
    +
    \ldots
    \Big\}
  \,.
\end{align}
Interestingly,
for any tensor field, this change is always equal to minus the
Lie derivative
under the flow of $\xi(x)$:
\begin{align}
  \label{eqn:lie}
  \change
  {T}^{\mu_1\mu_2\ldots}_{\nu_1\nu_2\ldots}
  =
  -
  \mathcal{L}_{\xi}\big[
    {T}^{\mu_1\mu_2\ldots}_{\nu_1\nu_2\ldots}
    \big]
  \,.
\end{align}
In fact, Hamilton uses this as the \emph{definition}
of the Lie derivative in his textbook~\cite{hamilton2018general}
(see Chapter 7.34 thereof).
This observation has motivated the extensive use of Lie derivatives
in many recent studies of the energy-momentum
tensor~\cite{Bilyalov:1992fd,Bilyalov:1996pe,GamboaSaravi:2002vos,GamboaSaravi:2003aq,Zhang:2004jb,Helfer:2016zvl,Kim:2024ewt}.

To make this less abstract, several concrete examples of
Eq.~(\ref{eqn:delta:linear}) are:
\begin{align}
  \label{eqn:examples}
  \begin{split}
    \change \phi
    =
    -\xi^\lambda \partial_\lambda \phi
    \qquad&:\qquad
    \text{\textbf{scalar field}}
    \\
    \change V^\mu
    =
    -\xi^\lambda \partial_\lambda V^\mu
    +
    (\partial_\lambda \xi^\mu) V^\lambda
    \qquad&:\qquad
    \text{\textbf{contravariant vector field}}
    \\
    \change A_\mu
    =
    -\xi^\lambda \partial_\lambda A_\mu
    -
    (\partial_\mu \xi^\lambda) A_\lambda
    \qquad&:\qquad
    \text{\textbf{covariant vector field}}
    \\
    \change F_{\mu\nu}
    =
    -\xi^\lambda \partial_\lambda F_{\mu\nu}
    -
    (\partial_\mu \xi^\lambda) F_{\lambda\nu}
    -
    (\partial_\nu \xi^\lambda) F_{\mu\lambda}
    \qquad&:\qquad
    \text{\textbf{rank-2 covariant tensor field}}
  \end{split}
\end{align}
If the metric tensor is considered a dynamical field,
then it should transform according to the rule for rank-2
covariant tensor fields.
However, the theories I will apply this method to are
quantum electrodynamics and chromodynamics in flat spacetime,
where the metric tensor is not considered a dynamical field.
Under an actual reparametrization of spacetime,
the metric should of course transform anyway,
but the transformation being considered is only a transformation
of the matter fields---not of spacetime itself.
The Minkowski metric $\eta_{\mu\nu}$
and its inverse $\eta^{\mu\nu}$ are thus left unchanged.

So far, I have addressed how tensor-valued matter fields transform
under a local translation.
It is also necessary to consider the transformation rules for
derivatives of the matter fields,
and transformation rules for spinor fields.
The first can potentially introduce complications because
coordinate derivatives of tensors are not necessarily tensors
under general coordinate transformations,
and therefore do not generically transform like tensors under
local translations either.
The second matter is subtle,
but ultimately spinor fields transform like scalar fields
under general coordinate transformations,
and therefore also transform this way under local translations.
I will address each of these matters in turn.


\subsection{Derivatives of tensor fields}

Aside from spinor fields (to be addressed below),
the fields appearing in the QED and QCD Lagrangians
are scalar fields and rank-1 covariant vector fields.

Derivatives of scalar fields do transform like tensors---specifically
like covariant vector fields---under general coordinate transformations.
To see this,
it is helpful to note that $\change$ and $\partial_\mu$ commute,
since differentiation distributes over addition.
Thus:
\begin{align}
  \change(\partial_\mu\phi)
  =
  \partial_\mu(\change\phi)
  =
  -
  \xi^\lambda \partial_\lambda
  \partial_\mu\phi
  -
  (\partial_\mu \xi^\lambda)
  \partial_\lambda \phi
  \,,
\end{align}
agreeing with the covariant vector field rule
in Eq.~(\ref{eqn:examples}).

By contrast, $\partial_\mu A_\nu$ does not transform like a rank-two
covariant tensor field.
Again using commutativity of $\change$ and $\partial_\mu$ gives:
\begin{align}
  \label{eqn:pseudotensor}
  \change(\partial_\mu A_\nu)
  =
  \partial_\mu(\change A_\nu)
  =
  -
  \xi^\lambda \partial_\lambda \partial_\mu A_\nu
  -
  (\partial_\mu \xi^\lambda) \partial_\lambda A_\nu
  -
  (\partial_\nu \xi^\lambda) \partial_\mu A_\lambda
  -
  (\partial_\mu \partial_\nu \xi^\lambda) A_\lambda
  \,.
\end{align}
The first three terms resemble the rank-2 covariant tensor field rule
in Eq.~(\ref{eqn:examples}),
but there is an extra term with two derivatives of $\xi^\lambda$.
This occurs because $\partial_\mu A_\nu$ is not a proper tensor;
normally, one must construct the covariant derivative
$\mathcal{D}_\mu A_\nu$, which would in fact transform like a rank-2 tensor.

On the other hand,
the antisymmetric combination
$\partial_{[\mu} A_{\nu]} = \partial_\mu A_\nu - \partial_\nu A_\mu$,
does transform like a proper rank-2 tensor.
This can be seen easily by antisymmetrizing Eq.~(\ref{eqn:pseudotensor}):
\begin{align}
  \change(\partial_{[\mu} A_{\nu]})
  =
  -
  \xi^\lambda \partial_\lambda \partial_{[\mu} A_{\nu]}
  -
  (\partial_\mu \xi^\lambda) \partial_{[\lambda} A_{\nu]}
  -
  (\partial_\nu \xi^\lambda) \partial_{[\mu} A_{\lambda]}
  \,.
\end{align}
Since $\partial_\mu A_\nu$ always appears in the QED and QCD Lagrangians
through this antisymmetric combination,
the ``extra'' term in Eq.~(\ref{eqn:pseudotensor})
can be dropped without affecting the result.

In fact, this can be shown explicitly.
Because the Lagrangian depends only on $\partial_\mu A_\nu$ through
its antisymmetrization, we have:
\begin{align*}
  \frac{\partial\Lag}{\partial(\partial_\mu A_\nu)}
  =
  -
  \frac{\partial\Lag}{\partial(\partial_\nu A_\mu)}
  \,,
\end{align*}
and accordingly
the relevant quantity appearing in Noether's second theorem can be rewritten:
\begin{align*}
  \frac{\partial\Lag}{\partial(\partial_\mu A_\nu)}
  \change(\partial_\mu A_\nu)
  =
  -
  \frac{\partial\Lag}{\partial(\partial_\nu A_\mu)}
  \change(\partial_\mu A_\nu)
  =
  \frac{1}{2}
  \frac{\partial\Lag}{\partial(\partial_\mu A_\nu)}
  \change(\partial_{[\mu} A_{\nu]})
  \,,
\end{align*}
meaning the ``extra'' term in Eq.~(\ref{eqn:pseudotensor})
is guaranteed to drop out.


\subsection{Local translations of spinor fields}

The main pretense of Ref.~\cite{Freese:2021jqs}
was that Eq.~(\ref{eqn:lie}) would generalize to spinors---i.e.,
that under a local translation, a spinor field $\psi$ would transform as
$\change \psi = -\mathcal{L}_\xi[\psi]$,
with the spinor Lie derivative having been given by Kosmann~\cite{Kosmann:1966}.
However, that pretense is questionable;
in the standard treatment,
spinor fields actually transform like scalars under local translations.

The underlying issue is that there is no finite linear
spinor representation of the group of general coordinate
transformations.
Classic proofs were given by Weyl~\cite{Weyl:1929fm}
and Cartan~\cite{Cartan:1966},
and I give an elementary proof in Appendix~\ref{sec:proof}
(along with some caveats about the existence of infinite and nonlinear spinor representations).
It is in fact exactly for this reason that the standard method of incorporating
spinors into theories with curved spacetime is the tetrad formalism.
In-depth expositions of this framework can be found in
Appendix J of Carroll~\cite{Carroll:2004st},
Chapter 11 of Hamilton~\cite{hamilton2018general},
and Chapter 12 of Weinberg~\cite{Weinberg:1972kfs},
but I will give a quick overview.

In the tetrad formalism, an orthonormal frame with a basis
$\big\{ e_{0}, e_{1}, e_{2}, e_{3} \big\}$---called the
tetrad---is assigned to every point in spacetime.
A Latin letter from the beginning of the alphabet is usually used as an index
to signify the basis vector, e.g., $e_{a}$.
The components of the basis vectors, $\tensor{e}{_a^\mu}$,
are called the vierbein.
The orthonormality of the tetrad is imposed through the inner products:
\begin{align}
  \label{eqn:vb:metric}
  \begin{split}
    \eta^{ab}
    \tensor{e}{_a^\mu}
    \tensor{e}{_b^\nu}
    &=
    g^{\mu\nu}
    \\
    g_{\mu\nu}
    \tensor{e}{_a^\mu}
    \tensor{e}{_b^\nu}
    &=
    \eta_{ab}
    \,,
  \end{split}
\end{align}
where $g_{\mu\nu}$ is the spacetime metric and $\eta_{ab}$ is the Minkowski metric.
The vierbein transform like contravariant four-vectors
under general coordinate transformations,
with the transformation applied to the spacetime index:
\begin{align}
  \tensor{\tilde{e}}{_a^\mu}(\xp)
  =
  \frac{\partial \xp^\mu}{\partial x^\nu}
  \tensor{e}{_a^\nu}(x)
  \,.
\end{align}
This is a transformation just of the \emph{components} of $e_{a}$,
and not of $e_{a}$ as an abstract vector.
After this transformation,
the tetrad remains orthonormal and Eq.~(\ref{eqn:vb:metric})
remains satisfied with $\eta_{ab}$ unchanged.

With the tetrad in place, it is also possible to define transformations
that change the tetrad basis and maintain the orthonormality relations (\ref{eqn:vb:metric}).
This transformation can also vary from point-to-point.
Because they preserve the Minkowski metric $\eta_{ab}$,
they consist of local Lorentz transformations:
\begin{align}
  \tensor{\tilde{e}}{_a^\mu}(x)
  =
  \tensor{\varLambda}{_a^b}(x)
  \tensor{e}{_b^\mu}(x)
  \,.
\end{align}
Because the Lorentz group does have a spinor representation,
it makes sense for objects
to transform like spinors under local Lorentz transformations.

In essence,
spinors are introduced by building the Clifford algebra
in the tetrad frame.
The spinors are spinors with respect to local Lorentz transformations,
and the gamma matrices $\gamma^a$ are defined with tetrad indices
rather than spacetime indices.
The Dirac Lagrangian for instance is written:
\begin{align}
  \Lag
  =
  \frac{i}{2}
  \bar\psi
  \gamma^a
  \tensor{e}{_a^\mu}(x)
  (\partial_\mu \psi)
  -
  \frac{i}{2}
  (\partial_\mu \bar\psi)
  \gamma^a
  \tensor{e}{_a^\mu}(x)
  \psi
  -
  m \bar\psi \psi
  \,.
\end{align}
The field $\psi$ transforms like a spinor
with respect to local Lorentz transformations,
thus accommodating its spinorial character.
One can also use this to define gamma matrices with a spacetime index:
\begin{align}
  \gamma^\mu(x)
  \equiv
  \gamma^a
  \tensor{e}{_a^\mu}(x)
  \,,
\end{align}
which I will use in this paper for compactness of notation.

With respect to general coordinate transformations,
on the other hand,
the spinor field $\psi(x)$ transforms like a scalar field.
It accordingly must transform like a scalar field
under local translations as well.
Therefore, the rules for changes in the spinor field, conjugate spinor field,
and their derivatives under local translations are:
\begin{align}
  \label{eqn:examples:spinor}
  \begin{split}
    \change \psi
    &=
    -\xi^\lambda \partial_\lambda \psi
    \\
    \change \bar\psi
    &=
    -\xi^\lambda \partial_\lambda \bar\psi
    \\
    \change (\partial_\mu\psi)
    &=
    -\xi^\lambda \partial_\lambda \partial_\mu \psi
    -
    (\partial_\mu \xi^\lambda)
    (\partial_\lambda \psi)
    \\
    \change (\partial_\mu\bar\psi)
    &=
    -\xi^\lambda \partial_\lambda \partial_\mu \bar\psi
    -
    (\partial_\mu \xi^\lambda)
    (\partial_\lambda \bar\psi)
    \,,
  \end{split}
\end{align}
None of these are equal to minus the Lie derivative---contradicting
Ref.~\cite{Freese:2021jqs}.

Lastly, under a local translation of the matter fields,
the vierbein remains unchanged.
The reason for this is in Eq.~(\ref{eqn:vb:metric}).
The local translation acts only on the matter fields, leaving the metric alone.
If the vierbein is transformed, then by Eq.~(\ref{eqn:vb:metric})
the metric must be as well.


\subsection{When are local translations a symmetry?}

It is not immediately obvious that local translations should be
a symmetry of the action,
and in fact in general they are not.
A local translation is effectively a symmetry of the action only when
the Euler-Lagrange equations of motion are observed.
The conserved current derived by assuming they are a symmetry will thus
only be conserved for on-shell states.

To see the connection, simply look back at Eq.~(\ref{eqn:chain}),
use the fact that $\change$ and $\partial_\nu$ commute,
perform integration by parts, and drop the surface terms:
\begin{align}
  \int \d^4 x \,
  \sum_a
  \left\{
    \frac{\partial\Lag}{\partial\Psi_a}
    -
    \partial_\nu
    \left[
      \frac{\partial\Lag}{\partial(\partial_\nu\Psi_a)}
      \right]
    \right\}
  \change \Psi_a
  =
  0
  \,.
\end{align}
This is satisfied whenever the Euler-Lagrange equations are.

A caveat I should raise before proceeding is that,
in its original context,
Noether's second theorem was meant only to apply to
mathematically trivial symmetries of the action---that is,
transformations for which
$\change S = 0$
without assuming any physical equations of motion.
However, by combining the second theorem with equations of motion,
it is possible to derive additional corollaries.
(See Ref.~\cite{Brading:2000hc} for an in-depth discussion of this.)
If
$\change S = 0$
holds under a specific set of conditions,
then it follows that
Eq.~(\ref{eqn:theorem})
is true under the same conditions.
Applied to local translations,
this will entail a continuity equation for an EMT
which holds for on-shell states.
(For off-shell states, it would need to be generalized
by a Ward-Takahashi identity;
see Refs.~\cite{Brout:1966oea,Freese:2019bhb} for examples of such identities
for the canonical EMT.)


\section{The energy-momentum tensor}
\label{sec:emt}

With Noether's second theorem and local translations both clearly defined,
we can move on to obtaining the energy-momentum tensor.
This basically involves calculating the coefficients
$\mathscr{A}^\nu$ and $\mathscr{B}^{\mu\nu}$
defined in Eq.~(\ref{eqn:ABC})
when a local translation is performed,
and then plugging them into Noether's second theorem~(\ref{eqn:theorem}).
This will result in a conserved current,
which is identified as the energy-momentum tensor.

To be sure, it is not immediately clear that
Eq.~(\ref{eqn:theorem}) as written entails a conservation law.
Either $\mathscr{A}^\nu$ needs to vanish, or else be equal to a divergence.
In fact, the latter will occur: it turns out that
$\mathscr{A}^\nu = -\eta^{\mu\nu}\partial_\mu\Lag$
for any field theory.
To see this, note that the transformation rule for every tensor and spinor,
as well as their derivatives, contains a term of the form:
\begin{align}
  \begin{split}
    \change \Psi_a
    &=
    -\xi^\nu
    \partial_\nu \Psi_a
    +
    \big\{
      \text{linear in $\partial\xi$}
      \big\}
    \\
    \change (\partial_\mu \Psi_a)
    &=
    -\xi^\nu
    \partial_\nu (\partial_\mu \Psi_a)
    +
    \big\{
      \text{linear in $\partial\xi$}
      \big\}
    \,.
  \end{split}
\end{align}
The terms linear in $\partial\xi$, which I have not explicitly written,
are those contributing to
$\mathscr{B}^{\mu\nu}$.
Keeping only the terms contributing to $\mathscr{A}^\nu$ gives:
\begin{align}
  -
  \sum_a
  \left\{
    \frac{\partial\Lag}{\partial\Psi_a}
    \partial_\nu \Psi_a
    +
    \frac{\partial\Lag}{\partial(\partial_\mu\Psi_a)}
    \partial_\nu (\partial_\mu \Psi_a)
    \right\}
  \xi^\nu
  =
  \mathscr{A}^\nu \xi_\nu
  \,.
\end{align}
By the chain rule, and by the arbitrariness of $\xi^\nu$, it follows that:
\begin{align}
  \label{eqn:A}
  \mathscr{A}^\nu
  =
  -\eta^{\mu\nu}
  \partial_\mu \Lag
  \,.
\end{align}
Putting this into Noether's second theorem~(\ref{eqn:theorem}) entails that
\begin{align}
  \label{eqn:emt}
  T^{\mu\nu}
  =
  -
  \mathscr{B}^{\mu\nu}
  -
  \eta^{\mu\nu}
  \Lag
\end{align}
is a conserved current,
and a candidate for the energy-momentum tensor.

The calculation of $\mathscr{B}^{\mu\nu}$ remains.
This will depend on the theory in question,
and amounts to an exercise in bookkeeping.
Like in Ref.~\cite{Freese:2021jqs},
I will consider quantum electrodynamics (QED)
and quantum chromodynamics (QCD) in turn,
but this time using the corrected transformation rule
(\ref{eqn:examples:spinor}) for spinors.


\subsection{Quantum electrodynamics}
\label{sec:qed}

Let's consider quantum electrodynamics (QED) first.
Just as in Ref.~\cite{Freese:2021jqs},
I use the Gupta-Bleuler formalism for gauge-fixing~\cite{Gupta:1949rh,Bleuler:1950cy},
and introduce a non-zero photon mass $\mu$ for infrared
regulation~\cite{Proca:1936fbw,Stueckelberg:1938hvi}.
The QED Lagrangian takes the form:
\begin{align}
  \label{eqn:lag:qed}
  \Lag_{\mathrm{QED}}
  =
  \bar\psi
  \left(
  \frac{i}{2}
  \overleftrightarrow{\slashed{\D}}
  -
  m
  \right)
  \psi
  -
  \frac{1}{4} F_{\mu\nu} F^{\mu\nu}
  +
  \frac{1}{2} \mu^2 A_\mu A^\mu
  -
  \frac{\lambda}{2} (\partial_\mu A^\mu)^2
  \,.
\end{align}
In this context, the slashed two-sided derivative should be interpreted
not to act on the vierbein:
\begin{align}
  \bar\psi
  \overleftrightarrow{\slashed{\D}}
  \psi
  \equiv
  \bar\psi
  \gamma^a
  \tensor{e}{_a^\mu}
  (\D_\mu \psi)
  -
  (\D_\mu\bar\psi)
  \gamma^a
  \tensor{e}{_a^\mu}
  \psi
  \,,
\end{align}
although since we are working in flat spacetime
and not transforming the metric (or the vierbein),
this doesn't actually matter for our purposes.
The gauge-covariant derivative is, as usual:
\begin{align}
  \begin{split}
    \D_\mu \psi
    &=
    \partial_\mu \psi
    +
    i e A_\mu \psi
    \\
    \D_\mu \bar\psi
    &=
    \partial_\mu \bar\psi
    -
    i e A_\mu \bar\psi
    \,.
  \end{split}
\end{align}

To obtain the EMT through local translation,
we need to evaluate the left-hand side of
Eq.~(\ref{eqn:ABC}),
given the QED Lagrangian (\ref{eqn:lag:qed}),
and isolate the terms linear in $\partial_\mu \xi_\nu$.
This will give us the $\mathscr{B}^{\mu\nu}$ term needed to
construct the EMT via Eq.~(\ref{eqn:emt}).
To this end, let us define on a per-field basis:
\begin{align}
  \label{eqn:B:breakdown}
  \begin{split}
    \frac{\partial\Lag}{\partial\Psi_a}
    \change \Psi_a
    & \equiv
    \mathscr{A}^\nu\big[\Psi_a\big] \xi_\nu(x)
    +
    \mathscr{B}^{\mu\nu}\big[\Psi_a\big] \partial_\mu \xi_\nu(x)
    \\
    \frac{\partial\Lag}{\partial(\partial_\rho\Psi_a)}
    \change (\partial_\rho\Psi_a)
    & \equiv
    \mathscr{A}^\nu\big[\partial_\rho\Psi_a\big] \xi_\nu(x)
    +
    \mathscr{B}^{\mu\nu}\big[\partial_\rho\Psi_a\big] \partial_\mu \xi_\nu(x)
    \,,
  \end{split}
\end{align}
so that:
\begin{align}
  \begin{split}
    \mathscr{A}^\nu
    &=
    \sum_a
    \Big\{
      \mathscr{A}^\nu\big[\Psi_a\big]
      +
      \mathscr{A}^\nu\big[\partial_\rho\Psi_a\big]
      \Big\}
    \\
    \mathscr{B}^{\mu\nu}
    &=
    \sum_a
    \Big\{
      \mathscr{B}^{\mu\nu}\big[\Psi_a\big]
      +
      \mathscr{B}^{\mu\nu}\big[\partial_\rho\Psi_a\big]
      \Big\}
    \,.
  \end{split}
\end{align}
It's a matter or rote calculation to perform the relevant functional derivatives,
to substitute in the transformation rules of Eqs.~(\ref{eqn:examples})
and (\ref{eqn:examples:spinor}),
and then to pull out the terms linear in derivatives of $\xi(x)$.
Doing the rote calculations gives the following results:
\begin{align}
  \begin{split}
    \mathscr{B}^{\mu\nu}\big[\psi\big]
    &=
    0
    \\
    \mathscr{B}^{\mu\nu}\big[\partial_\rho \psi\big]
    &=
    -
    \frac{i}{2}
    \bar\psi
    \gamma^\mu
    (\partial^\nu \psi)
    \\
    \mathscr{B}^{\mu\nu}\big[\bar\psi\big]
    &=
    0
    \\
    \mathscr{B}^{\mu\nu}\big[\partial_\rho \bar\psi\big]
    &=
    \frac{i}{2}
    (\partial^\nu \bar\psi)
    \gamma^\mu
    \psi
    \\
    \mathscr{B}^{\mu\nu}\big[A_\rho\big]
    &=
    e\bar\psi \gamma^\mu A^\nu \psi
    -
    \mu^2 A^\mu A^\nu
    \\
    \mathscr{B}^{\mu\nu}\big[\partial_\rho A_\tau\big]
    &=
    -
    F^{\mu\rho}
    F_{\rho}^{\phantom{\rho}\nu}
    +
    \lambda
    (\partial^{\{\mu} A^{\nu\}})
    (\partial_\rho A^\rho)
    \,.
  \end{split}
\end{align}
Adding these pieces together, and using Eq.~(\ref{eqn:emt})
gives the following EMT:
\begin{align}
  \label{eqn:emt:qed}
  T^{\mu\nu}_{\mathrm{QED}}
  =
  \frac{i}{2}
  \bar\psi
  \gamma^\mu
  \overleftrightarrow{\D}^\nu
  \psi
  +
  \tensor{F}{^\mu^\rho} \tensor{F}{_\rho^\nu}
  +
  \mu A^\mu A^\nu
  -
  \lambda
  (\partial^{\{\mu} A^{\nu\}})
  (\partial_\rho A^\rho)
  -
  \eta^{\mu\nu} \Lag_{\mathrm{QED}}
  \,.
\end{align}
When the photon mass and gauge-fixing terms are removed,
this EMT is gauge-invariant, as expected.
However, in contrast to the result in Ref.~\cite{Freese:2021jqs},
it is not symmetric.
The piece
$
  \frac{i}{2}
  \bar\psi
  \gamma^\mu
  \overleftrightarrow{\D}^\nu
  \psi
$
involving the fermion field in particular is asymmetric.


\subsection{Quantum chromodynamics}
\label{sec:qcd}

Let's finally consider quantum chromodynamics (QCD).
I will use the Lagrangian given by
Kugo and Ojima~\cite{Kugo:1979gm}:
\begin{align}
  \label{eqn:lag:qcd}
  \Lag_{\mathrm{QCD}}
  =
  \sum_q
  \bar{q}
  \left(
  \frac{i}{2}
  \overleftrightarrow{\slashed{\D}}
  -
  m_q
  \right)
  q
  -
  \frac{1}{4} F^a_{\mu\nu} F_a^{\mu\nu}
  - (\partial_\mu B_a) A^\mu_a
  +
  \frac{\alpha_0}{2} B_a^2
  -
  i (\partial_\mu \bar{c}^a) (\D^\mu_{ab} c^b)
  \,,
\end{align}
where $a$ is an $\mathrm{SU}(3,\mathbb{C})$ color index
rather than a tetrad index.
Besides the quark fields $q$,
the gluon four-potential $A_\mu^a$,
and the gluon field strength tensor:
\begin{align}
  F_{\mu\nu}^a
  =
  \partial_\mu
  A_\nu^a
  -
  \partial_\nu
  A_\mu^a
  +
  g f_{abc} A_\mu^b A_\nu^c
  \,,
\end{align}
the Lagrangian (\ref{eqn:lag:qcd}) depends on
Lagrange multiplier fields $B_a$
(used for gauge fixing)
Faddeev-Popov ghosts $c_a$ and $\bar{c}_a$
(used to subtract contributions from unphysical gluon modes)~\cite{Faddeev:1967fc}.
The relevant representations of the gauge-covariant derivative are:
\begin{align}
  \begin{split}
    \D_\mu q
    &=
    \partial_\mu q
    -
    i g A^a_\mu T_a q
    \\
    \D_\mu \bar{q}
    &=
    \partial_\mu \bar{q}
    +
    i g \bar{q} A^a_\mu T_a
    \\
    \D_\mu^{ab} c^b
    &=
    \Big(
    \delta_{ab} \partial_\mu
    + g f_{acb} A_\mu^c
    \Big)
    c^b
    \,,
  \end{split}
\end{align}
$T_a$ are the generators of the $\mathfrak{su}(3,\mathbb{C})$
color algebra,
and $f_{abc}$ are the color algebra structure constants:
\begin{align}
  [T_a, T_b]
  =
  i f_{abc} T_c
  \,.
\end{align}

Before proceeding to obtain the EMT,
we should dwell on the issue of what symmetries QCD should observe.
The QCD Lagrangian with gauge-fixing terms is not gauge-invariant per se,
but it is invariant under the larger
Becchi-Rouet-Stora-Tyutin (BRST) transformation
group~\cite{Becchi:1974md,Becchi:1975nq,Tyutin:1975qk},
which confers upon QCD all the perks of gauge invariance
(such as renormalizability~\cite{Collins:1984xc}).
A BRST transformation changes the fields appearing in
in the QCD Lagrangian as
follows~\cite{Becchi:1974md,Becchi:1975nq,Tyutin:1975qk,Kugo:1979gm}:
\begin{align}
  \label{eqn:brst}
  \begin{split}
    \delta_{\mathrm{BRST}} A_\mu^a
    &=
    \lambda
    \D_\mu^{ab} c^b
    \\
    \delta_{\mathrm{BRST}} c^a
    &=
    -
    \frac{1}{2}
    \lambda
    \,
    g f_{abc} c_b c_c
    \\
    \delta_{\mathrm{BRST}} c^a
    &=
    i
    \lambda
    B^a
    \\
    \delta_{\mathrm{BRST}} B^a
    &=
    0
    \\
    \delta_{\mathrm{BRST}} q
    &=
    i\, T_a \lambda c_a q
    \,,
  \end{split}
\end{align}
where $\lambda$ is a Grassmann-number-valued parameter.
Since $c_a$ is also Grassmann-number-valued,
the quark and gluon fields transform as they would
under an infinitesimal gauge transformation:
\begin{align}
  \begin{split}
    \delta_{\mathrm{gauge}} q
    &=
    i\, T_a \theta_a q
    \\
    \delta_{\mathrm{gauge}} A_\mu^a
    &=
    \big(
    \delta_{ab}
    \partial_\mu
    +
    g f_{abc} A^c_\mu
    \big)
    \theta_b
    \,,
  \end{split}
\end{align}
with the product $\lambda c_a$ playing the role of $\theta_a$.
It is just a matter of rote calculation to show
that the QCD Lagrangian (\ref{eqn:lag:qcd})
is invariant under the BRST transformation
(\ref{eqn:brst}).

A full exposition of the BRST formalism can be found in
Kugo and Ojima~\cite{Kugo:1979gm}.
The transformation rules (\ref{eqn:brst}) are important here
because the QCD energy-momentum tensor should be invariant
under BRST transformations.

Let us move on to obtaining the EMT that follows from
the Lagrangian (\ref{eqn:lag:qcd})
and local translation invariance.
As in the QED case,
I'll step through individual fields' contributions
to the $\mathscr{B}^{\mu\nu}$ coefficient; see Eq.~(\ref{eqn:B:breakdown}).
Using the transformation rules (\ref{eqn:examples})
and (\ref{eqn:examples:spinor}),
the non-trivial $\mathscr{B}^{\mu\nu}$ coefficients evaluate to:
\begin{align}
  \begin{split}
    \mathscr{B}^{\mu\nu}\big[\partial_\rho q\big]
    &=
    -
    \frac{i}{2}
    \bar{q}
    \gamma^\mu
    (\partial^\nu q)
    \\
    \mathscr{B}^{\mu\nu}\big[\partial_\rho \bar{q}\big]
    &=
    \frac{i}{2}
    (\partial^\nu \bar{q})
    \gamma^\mu
    q
    \\
    \mathscr{B}^{\mu\nu}\big[A^a_\rho\big]
    &=
    -
    \sum_q
    g \bar q \gamma^\mu A^\nu T_a q
    -
    g f_{abc} F_b^{\mu\rho} A^c_\rho A^\nu_a
    +
    i g f_{abc} (\partial_\mu \bar{c}^c) A^\nu_a c^b
    +
    (\partial^\mu B_a) A^\nu_a
    \\
    \mathscr{B}^{\mu\nu}\big[\partial_\rho A^a_\tau\big]
    &=
    -
    F^{\mu\rho}_a
    ( \partial_\rho A^\nu_a - \partial^\nu A_\rho^a )
    \\
    \mathscr{B}^{\mu\nu}\big[\partial_\rho B_a\big]
    &=
    A^\mu_a (\partial^\nu B_a)
    \\
    \mathscr{B}^{\mu\nu}\big[\partial_\rho c_a\big]
    &=
    i (\partial^\nu \bar{c}^a) (\partial^\mu c^a)
    \\
    \mathscr{B}^{\mu\nu}\big[\partial_\rho \bar{c}_a\big]
    &=
    i (\partial^\nu \bar{c}^a)(\D^\mu_{ab} c^b)
    \,,
  \end{split}
\end{align}
with the remaining coefficients being zero.
Adding these together, and using Eq.~(\ref{eqn:emt}),
gives the following energy-momentum tensor:
\begin{align}
  \label{eqn:emt:qcd}
  T^{\mu\nu}_{\mathrm{QCD}}
  =
  \sum_q
  \frac{i}{2}
  \bar{q}
  \gamma^{\mu}
  \overleftrightarrow{\D}^{\nu}
  q
  +
  F_a^{\mu\rho} F_{\rho}^{a\,\nu}
  -
  A_a^{\{\mu} \partial^{\nu\}}_{\phantom{a}} B_a
  -
  i (\D^{\{\mu} c) (\partial^{\nu\}} \bar{c})
  -
  \eta^{\mu\nu} \Lag_{\mathrm{QCD}}
  \,.
\end{align}
As in the QED case, the result is asymmetric---specifically
in the terms
$
  \sum_q
  \frac{i}{2}
  \bar{q}
  \gamma^{\mu}
  \overleftrightarrow{\D}^{\nu}
  q
$
involving the quark fields.
This EMT is BRST invariant
(which can be checked by performing the transformation
(\ref{eqn:brst})
and working out the algebra),
and when restricted to physical states
(which are annihilated by $B_a$, $c_a$ and $\bar{c}_a$)
is gauge-invariant.
Since it observes the symmetries of QCD,
there is no a priori reason to reject it as unphysical,
even if it is asymmetric.

The EMT in Eq.~(\ref{eqn:emt:qcd}) is identical to the
``gauge-invariant canonical''
or
``gauge-invariant kinetic''\footnote{
  The total EMT is identical in both cases;
  the ``canonical'' and ``kinetic'' differ
  only in how they're broken down into quark and gluon pieces.
}
EMT identified by Leader and Lorc\'e~\cite{Leader:2013jra},
which is also the asymmetric EMT in common use in the hadron physics literature.


\section{Summary and outlook}
\label{sec:end}

In this work, I revisited the derivation of the energy momentum tensors
of quantum electrodynamics and quantum chromodynamics
through Noether's second theorem,
with local translations of the matter fields as the relevant symmetry.
I previously gave a similar derivation
in Ref.~\cite{Freese:2021jqs},
obtaining the symmetric Belinfante EMT in both theories,
but had committed a grievous error
in the transformation of the spinor fields.
There I had, in effect,
assumed the existence of a finite linear spinor representation
of the general coordinate transformation group.
However, this assumption was wrong,
and the transformation rule used
in Ref.~\cite{Freese:2021jqs}
was unsound.

The correct transformation rule for spinor fields
under local translations is given by Eq.~(\ref{eqn:examples:spinor}).
Spinors effectively transform like scalar fields under local translations,
just as they transform like scalar fields under general coordinate transformations.
(The spinorial character is instead incorporated by
the field's transformation properties
under a change of local orthonormal frame.)
Using the corrected transformation rule,
the conclusions
of Ref.~\cite{Freese:2021jqs}
are altered:
instead, the gauge-invariant \emph{asymmetric} EMT is obtained in both theories.
For QCD in particular, the result---given in Eq.~(\ref{eqn:emt:qcd})---agrees
with the ``gauge-invariant canonical'' or ``gauge-invariant kinetic'' EMT
of Leader and Lorc\'e~\cite{Leader:2013jra}.

Ultimately, the symmetric and asymmetric EMTs
are both conserved quantities.
What I have shown by correcting
Ref.~\cite{Freese:2021jqs}
is that the asymmetric EMT is
the conserved current associated with local translation symmetry.
This does not, however, directly address the question of
whether the EMT in nature is symmetric or asymmetric.
The symmetric EMT is instead obtained under other definitions;
as observed by Kugo and Ojima~\cite{Kugo:1979gm},
differentiating the QCD action with respect to the vierbein
gives the symmetric EMT---at least assuming the
Levi-Civita connection\footnote{
  In curved spacetime, the derivatives of $q$ and $\bar{q}$ are replaced
  by covariant derivatives in the QCD action, and these depend on the spin connection.
  The resulting EMT depends on how the spin connection is defined.
  One recovers the symmetric EMT if the Levi-Civita connection is used,
  which in effect amounts to assuming there's no spacetime torsion.
  A different EMT may be obtained assuming a different connection.
  In fact, Iosifidis \textsl{et al.}~\cite{Iosifidis:2025sjx}
  show that for metric-affine gravity, in which the spin connection
  is considered independent of the vierbein,
  this procedure instead gives the canonical EMT.
}.
The answer to this question must instead come from experiment.

It is currently unclear whether an antisymmetric component of the EMT
could be probed in a fixed-target or collider experiment.
In the realm of astrophysics and cosmology,
where the EMT is the source of gravitation,
an asymmetric EMT would necessarily produce spacetime torsion;
see Refs.~\cite{Hehl:1976kj,Hammond:2002rm,Shapiro:2001rz}
for reviews on torsion theories of gravity.

Einstein-Cartan theory~\cite{cartan1922generalisation}
(also rediscovered by
Kibble~\cite{Kibble:1961ba}
and Sciama~\cite{Sciama:1964jqa})
accommodates spacetime torsion
by minimally coupling the torsion tensor to the matter fields.
However, torsion does not
propagate outside matter in this theory,
which makes empirical confirmation of spacetime torsion a tricky matter.
However, spacetime torsion can still result in effective spin-spin
interactions, which could modify the equation of state in neutron stars.
Jockel \textsl{et al.}~\cite{Jockel:2024fps}
have shown
the presence of spacetime torsion
can result in more compact neutron stars
with greater central densities.
It may be possible to test this prediction in the near future
through gravitational wave
asteroseismology~\cite{Andersson:1997rn,Benhar:2004xg,Tsui:2005zf,Sotani:2012qc}
of neutron star-black hole mergers.

Another potential difficulty
is that naively performing minimal substitution for all the fields---including
gauge fields---results in the canonical EMT beng the source of the Einstein tensor,
and the field equations thus being gauge-dependent.
In particular, replacing $\partial_\mu$ by the covariant derivative $\nabla_\mu$
in the field strength tensor
gives
$\nabla_\mu A_\nu - \nabla_\nu A_\mu \neq \partial_\mu A_\nu - \partial_\nu A_\mu$
in the presence of torsion,
with the extra piece being gauge-dependent~\cite{Shapiro:2001rz}.
On the other hand, as many authors have
observed~\cite{Weinberg:1995mt,hehl2003foundations,Lorce:2012rr,Hehl:2014eja},
$A_\mu$ is not actually a four-vector,
but is \emph{defined} so that its exterior derivative---which involves
only the coordinate derivative, and not the covariant derivative---gives
the field strength tensor $F_{\mu\nu}$.
Thus, minimal substitution should be performed only for non-gauge fields\footnote{
  In fact, Shapiro~\cite{Shapiro:2001rz}
  finds that not only can Abelian gauge fields not couple minimally to torsion
  without breaking gauge invariance,
  but non-Abelian gauge fields cannot couple to torsion at all.
}.
From this perspective, the QCD EMT of Eq.~(\ref{eqn:emt:qcd})
seems quite reasonable as a source for torsion gravity.

In any case,
the question of whether the EMT in nature is symmetric or asymmetric
remains open.
Since the asymmetric EMT differs only by the addition of
an antisymmetric part that is parametrized by one
mechanical form factor~\cite{Lorce:2018egm},
it is perhaps prudent to consider the asymmetric EMT in theoretical studies
for full generality.
The antisymmetric form factor can simply be set to zero
in the parametrization if one wants to consider the symmetric EMT.


\begin{acknowledgments}
  I gratefully acknowledge valuable discussions with
  Friedrich Hehl, Damianos Iosifidis, Brian Pitts,
  Gabriel Santiago and Alan Sosa,
  and also thank Alan Sosa for a figure I based Fig.~\ref{fig:cover} on.
  The research in this work received inspiration from the goals of the
  Quark Gluon Tomography Topical Collaboration of the U.S.\ Department of Energy.
  This work was supported by
  the Center for Nuclear Femtography,
  operated by the Southeastern Universities Research Association
  in Washington, D.C.\ under an appropriation from the Commonwealth of Virginia;
  by the DOE contract No.~DE-AC05-06OR23177,
  under which Jefferson Science Associates, LLC operates Jefferson Lab;
  and by the Scientific Discovery through Advanced Computing (SciDAC) award
  \textsl{Femtoscale Imaging of Nuclei using Exascale Platforms}.
\end{acknowledgments}


\appendix

\section{Spinors and general coordinate transformations}
\label{sec:proof}

In this Appendix,
I give an elementary proof that
there is no finite linear spinor representation
of the group of general coordinate transformations.
Classic proofs were already given by Weyl~\cite{Weyl:1929fm}
and Cartan~\cite{Cartan:1966}.
The proof herein aims to be as elementary and broadly accessible as possible,
in effect reducing the problem to that of classifying spin representations
through ladder operators.


\subsection{The proof}

A general coordinate transformation is effectively
a matrix from the general linear group
$\mathrm{GL}(4,\mathbb{R})$---that is,
the group of all $4\times4$ real-valued matrices with non-zero
determinant---assigned to every point in spacetime.
The particular matrix
$\tensor{\Lambda}{^\mu_\nu}(x)\in\mathrm{GL}(4,\mathbb{R})$
is given by the transformation rule for contravariant vector fields:
\begin{align}
  V'^\mu(x')
  =
  \Lambda^{\mu}_{\phantom{\mu}\nu}(x)
  V^\nu(x)
  =
  \frac{\partial x'^\mu}{\partial x^\nu}
  V^\nu(x)
  \,.
\end{align}
A spinor representation of this group would need to be
a two-valued representation.
More formally, we need another transformation group $\mathrm{ML}(4,\mathbb{R})$
that has two transformations in $\mathrm{ML}(4,\mathbb{R})$ for every one transformation in
$\mathrm{GL}(4,\mathbb{R})$---that is, a double-cover---and
a map
\begin{align}
  \rho : \mathrm{ML}(4,\mathbb{R}) \rightarrow \mathrm{GL}(4,\mathbb{R})
\end{align}
that preserves the group structure
(so is a Lie group homomorphism).
The double cover would generalize the peculiar property of spinors
that they only return to their original state upon two full rotations.

The issue---and the matter to be proved---is there is no matrix group
that double-covers
$\mathrm{GL}(4,\mathbb{R})$.
To be sure, a double cover of
$\mathrm{GL}(4,\mathbb{R})$
does actually exist as a Lie group:
it is called the metalinear group---hence
the name $\mathrm{ML}(4,\mathbb{R})$ for the double-cover.
However, the metalinear group is not a matrix group,
which would make its use in transformation laws problematic.
(I will discuss what this means briefly after the proof.)

\begin{figure}
  \includegraphics[width=0.5\textwidth]{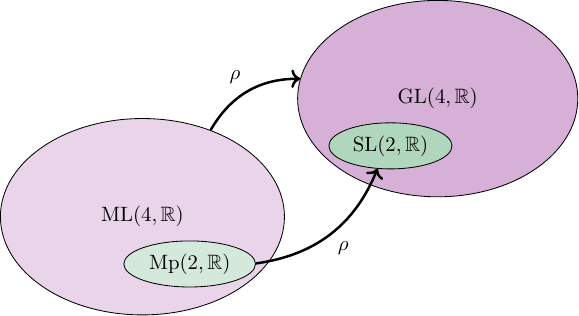}
  \caption{
    Depiction of a Lie group homomorphism
    $\rho : \mathrm{ML}(4,\mathbb{R}) \rightarrow \mathrm{GL}(4,\mathbb{R})$,
    which maps the double cover of
    $\mathrm{GL}(4,\mathbb{R})$
    (here marked $\mathrm{ML}(4,\mathbb{R})$)
    onto $\mathrm{GL}(4,\mathbb{R})$.
    Since
    $\mathrm{SL}(2,\mathbb{R})$
    is a subgroup of
    $\mathrm{GL}(4,\mathbb{R})$,
    its double cover
    (here marked $\mathrm{Mp}(2,\mathbb{R})$)
    must be a subgroup of
    $\mathrm{ML}(4,\mathbb{R})$---and
    $\rho$ must likewise map
    $\mathrm{Mp}(2,\mathbb{R})$
    onto
    $\mathrm{SL}(2,\mathbb{R})$.
    A faithful matrix representation of
    $\mathrm{ML}(4,\mathbb{R})$
    can only exist if a there is a faithful matrix representation of its subgroup
    $\mathrm{Mp}(2,\mathbb{R})$.
  }
  \label{fig:cover}
\end{figure}

I will start the proof following the observation by Cartan
(Ref.~\cite{Cartan:1966}, Section 177)
that since
$\mathrm{SL}(2,\mathbb{R})$---the special linear group of $2\times2$ matrices
with real components and determinant $1$---is a subgroup of
$\mathrm{GL}(4,\mathbb{R})$,
a hypothetical spinor representation of
$\mathrm{GL}(4,\mathbb{R})$
must contain a spinor representation of
$\mathrm{SL}(2,\mathbb{R})$
as a subgroup;
see Fig.~\ref{fig:cover}.
It thus suffices to prove that there is no matrix group
that is a double cover of
$\mathrm{SL}(2,\mathbb{R})$.

To do this,
let us consider matrix representations of the Lie algebra
$\mathfrak{sl}(2,\mathbb{R})$ that generates
$\mathrm{SL}(2,\mathbb{R})$.
A matrix $g\in\mathfrak{sl}(2,\mathbb{R})$ is converted to a matrix
$G\in\mathrm{SL}(2,\mathbb{R})$ through exponentiation:
\begin{align}
  G
  =
  \e^{g}
  \,,
\end{align}
and thus $\mathrm{det}(G) = 1$ implies $\mathrm{Tr}(g) = 2\pi i n$,
for some integer $n$.
$G$ being real-valued requires that $g$ is real-valued,
so $n=0$.
A basis for the Lie algebra
$\mathfrak{sl}(2,\mathbb{R})$
is thus:
\begin{align}
  \label{eqn:sl2R:def}
  \begin{split}
    J_z
    &=
    \frac{1}{2}
    \sigma_z
    =
    \frac{1}{2}
    \begin{bmatrix*}[r]
      1 & 0 \\
      0 & -1
    \end{bmatrix*}
    \\
    J_+
    &=
    \frac{1}{2}
    \big(
    \sigma_x
    +
    i \sigma_y
    \big)
    =
    \begin{bmatrix}
      0 & 1 \\
      0 & 0
    \end{bmatrix}
    \\
    J_-
    &=
    \frac{1}{2}
    \big(
    \sigma_x
    -
    i \sigma_y
    \big)
    =
    \begin{bmatrix}
      0 & 0 \\
      1 & 0
    \end{bmatrix}
    \,,
  \end{split}
\end{align}
where $\sigma_{x,y,z}$ are the usual Pauli matrices.
The commutation relations defining the abstract Lie algebra are:
\begin{align}
  \label{eqn:sl2R}
  [J_z, J_+]
  =
  J_+
  \qquad
  [J_z, J_-]
  =
  -
  J_-
  \qquad
  [J_+, J_-]
  =
  2
  J_z
  \,.
\end{align}
This is the familiar algebra of angular momentum ladder operators,
commonly found in quantum mechanics textbooks
(see for instance
Chapter 14 of Bohm~\cite{bohm1989quantum},
Chapter VI of Cohen-Tannoudji, Diu and Laloë~\cite{cohen2019quantum},
or
Chapter 3 of Sakurai and Napolitano~\cite{Sakurai:2011zz}).
These commutation relations need to be satisfied by any representation of
$\mathrm{SL}(2,\mathbb{R})$---including multi-valued representations
(such as a double cover).
If a matrix group that double-covers
$\mathrm{SL}(2,\mathbb{R})$
exists,
then it is generated by a matrix representation of
the commutation relations~(\ref{eqn:sl2R}).

The goal is now to prove that
there is no matrix representation of
$\mathfrak{sl}(2,\mathbb{R})$
that exponentiates to a double cover of
$\mathrm{SL}(2,\mathbb{R})$.
This is done by classifying all the matrix representations of
$\mathfrak{sl}(2,\mathbb{R})$,
and showing that none of them generates
a double cover.
Here, the fact that Eq.~(\ref{eqn:sl2R}) is the algebra of angular
momentum ladder operators is helpful.
The matrix representations can be classified by the Casimir invariant:
\begin{align}
  J^2
  \equiv
  J_z^2
  +
  \frac{1}{2}
  \big(
  J_+ J_-
  +
  J_- J_+
  \big)
  \,,
\end{align}
which commutes with all the generators.
Irreducible matrix representations of
$\mathfrak{sl}(2,\mathbb{R})$
are classified by the value of $J^2$,
and reducible representations can be written as the direct sum
of irreducible representations.
For instance, for
$\mathrm{SL}(2,\mathbb{R})$
itself,
$J^2 = \frac{3}{4}$ and $j=\frac{1}{2}$,
which can be worked out directly from Eq.~(\ref{eqn:sl2R:def}).

Next, given an irreducible matrix representation of
$\big\{J_z, J_+, J_-\big\}$,
it is possible to build matrices from complex linear
combinations of these generators.
Such matrices include:
\begin{align}
  \label{eqn:Jxy}
  J_x
  =
  \frac{J_+ + J_-}{2}
  \qquad
  \qquad
  J_y
  =
  \frac{J_+ - J_-}{2i}
  \,,
\end{align}
which together with $J_z$ satisfy the algebra of
$\mathfrak{su}(2,\mathbb{C})$:
\begin{align}
  [J_a, J_b]
  =
  i \epsilon_{abc} J_c
  \,.
\end{align}
To be sure, $J_x$ and $J_y$ are not in 
$\mathfrak{sl}(2,\mathbb{R})$,
nor are $J_+$ and $J_-$ in 
$\mathfrak{su}(2,\mathbb{C})$;
these algebras are closed only under real linear combinations
of the generators.
However, by considering these complex linear combinations,
we can see that the present representations of both algebras
have the same Casimir invariant:
\begin{align}
  J^2
  =
  J_z^2
  +
  \frac{1}{2}
  \big(
  J_+ J_-
  +
  J_- J_+
  \big)
  =
  J_x^2 + J_y^2 + J_z^2
  \,.
\end{align}
Thus, there is a one-to-one correspondence
between irreducible matrix representations
of both algebras.
The irreducible matrix representations of
$\mathfrak{su}(2,\mathbb{C})$ are well-known;
they are classified by
\begin{align}
  J^2
  =
  j(j+1)
  \,,
\end{align}
where $j$ can take on non-negative integer or half-integer values:
$
  j
  \in
  \big\{
    0,
    \tfrac{1}{2},
    1,
    \tfrac{3}{2},
    2,
    \ldots
    \big\}
$.
Additionally, $J_z$ is diagonalizable and can take on the
$2j+1$ eigenvalues from
$\big\{-j, -j+1, \ldots, j-1, j \big\}$,
so the matrices in question are $(2j+1)\times(2j+1)$ square matrices.

Now, let us consider an irreducible matrix representation $A(j)$ of
$\mathrm{SL}(2,\mathbb{R})$
with a definite Casimiar invariant $J^2 = j(j+1) \neq 0$
(since $j=0$ is the trivial representation),
and let $J_{+,-,z}^{(j)}$ be the generators in this representation.
$A(j)$ contains an Abelian subgroup consisting of matrices of the form
\begin{align}
  R(j,\phi)
  =
  \exp\left\{
    \tfrac{1}{2}
    \big( J_+^{(j)} - J_-^{(j)} \big)
    \phi
    \right\}
  =
  \exp\left\{
    i J_y^{(j)} \phi
    \right\}
  \,,
\end{align}
where
$J_y^{(j)}$ is a Hermitian matrix with the same eigenvalue spectrum as $J_z^{(j)}$\footnote{
  Although $J_y^{(j)}$ is not in the algebra of $A(j)$ itself,
  it does exist as a matrix in the complexification of the algebra,
  i.e., in the space of matrices spanned by complex linear combinations
  of the generators of $A(j)$.
}.
Let $m_j$ be the smallest positive eigenvalue of $J_y^{(j)}$,
which is $1$ for integer $j>0$
and $\frac{1}{2}$ for half-integer $j$.
There is a vector $|m_j\rangle$ for which
$J_y^{(j)} |m_j\rangle = m_j |m_j\rangle$, and thus for which:
\begin{align}
  R(j,\phi)
  |m_j \rangle
  =
  \e^{i m_j \phi}
  |m_j \rangle
  \,.
\end{align}
The Abelian subgroup of $A(j)$ consists of all unique matrices
in the $|m_j\rangle$ basis.
Thus, if $j$ is an integer,
then $\phi\in[0,2\pi)$ exhausts the subgroup;
whereas if $j$ is a half-integer then $\phi\in[0,4\pi)$
exhausts the subgroup.

Now, for another representation $A(\tilde{\textsl{\j}})$ to be a double cover of $A(j)$,
there should be a Lie group homomorphism
\begin{align}
  \rho : A(\tilde{\textsl{\j}}) \rightarrow A(j)
\end{align}
that maps two matrices from $A(\tilde{\textsl{\j}})$ to each matrix in $A(j)$.
This requires mapping two matrices to each distinct $R(j,\phi)$,
doubling the domain of $\phi$---to $[0,4\pi)$ for integer $j$,
or to $[0,8\pi)$ for half-integer $j$---which in turn
requires the smallest $\widetilde{m}_{\tilde{\textsl{\j}}}$ be half the smallest $m_j$.
For half-integer $j$, this would require $\widetilde{m}_{\tilde{\textsl{\j}}}=\frac{1}{4}$,
but there is no matrix representation of
$\mathrm{SL}(2,\mathbb{R})$
for which this is true.
Since $j=\frac{1}{2}$ for
$\mathrm{SL}(2,\mathbb{R})$
itself,
it follows that no matrix group exists
that is a double cover of
$\mathrm{SL}(2,\mathbb{R})$---and
therefore that there is no matrix group that double-covers
$\mathrm{GL}(4,\mathbb{R})$.
This completes the proof.


\subsection{Further discussion}

A caveat worth mentioning is that while there is no
\emph{matrix group} that double-covers
$\mathrm{GL}(4,\mathbb{R})$,
there is a Lie group that does:
the metalinear group,
$\mathrm{ML}(4,\mathbb{R})$---hence the label in Fig.~\ref{fig:cover}.
Its construction is quite technical,
and beyond the scope of this appendix;
see Chapter 7 of \cite{bates1997lectures}
or Section 3.2 of Ref.~\cite{bongers2014geometric}
for further details.

Similarly, while there is no matrix group that double-covers
$\mathrm{SL}(2,\mathbb{R})$,
there is a Lie group that does:
the metaplectic group
$\mathrm{Mp}(2,\mathbb{R})$---hence the label in Fig.~\ref{fig:cover}.
(Notably, the metaplectic group is not the \emph{universal} cover of
$\mathrm{SL}(2,\mathbb{R})$,
just the double cover.
The universal cover of
$\mathrm{SL}(2,\mathbb{R})$
has no special name,
and is simply denoted
$\widetilde{\mathrm{SL}}(2,\mathbb{R})$;
see Refs.~\cite{pukanszky1964plancherel,Kitaev:2017hnr} for details,
and Section 86 of Cartan's book~\cite{Cartan:1966} for a construction.)
A friendly introduction to metaplectic groups,
including a short proof that
$\mathrm{Mp}(2,\mathbb{R})$
has no faithful matrix representation,
can be found in Ref.~\cite{Weismann:2023}.

Rather than by matrices,
metaplectic and metalinear groups can be represented by unitary operators
on (infinite-dimensional) Hilbert space~\cite{weil1964certains,pukanszky1964plancherel,Weismann:2023}.
One may wonder why spinor fields can't simply be Hilbert space vectors, then,
which transform under these unitary operations
when a general coordinate transformation is performed.
The issue with this is that a spinor field $\psi(x)$
is a spinor-valued field---it assigns a spinor to every point in spacetime,
each of which would need to transform separately under a different operator from
$\mathrm{ML}(4,\mathbb{R})$.
In other words,
this construction would involve assigning a vector from
an infinite-dimensional Hilbert space to every point in spacetime.
(Note that this is quite distinct from a wave function,
which is a number-valued function of space and time
corresponding to a single Hilbert space vector.)
A spinor field of this kind would have infinitely many degrees of freedom
and not look anything like the four-component spinors of quantum chromodynamics.

Another way of circumventing the lack of faithful matrix representations
may be to realize finite but non-linear representations of
$\mathrm{ML}(4,\mathbb{R})$.
The approaches of
Ogievetsky and Polubarinov~\cite{Ogievetsky:1965ii}
and of Bilyalov~\cite{Bilyalov:1992fd}
seem to be along these lines.
(Also see Pitts~\cite{Pitts:2011jv} for a detailed overview.)
In these approaches,
coordinate transformations act on a pair of objects
rather than the spinor field $\psi(x)$ alone.
In the Ogievetsky-Polubarinov formalism,
this pair is $(\psi(x),r_{\mu\nu}(x))$,
where $r_{\mu\nu}(x)$ is a ``square root of the metric.''
(See Refs.~\cite{Ogievetsky:1965ii,Bilyalov:1992fd}
for details on what this means.)
However, the transformations constructed by Ogievetsky and Polubarinov
do not form a faithful representation
of the full double-cover of $\mathrm{GL}(4,\mathbb{R})$,
because they exclude transformations that allow coordinate reordering.
In fact, this formalism cannot accommodate transformations from
instant form to light front coordinates---which
is a dealbreaker for hadron physics,
where light front coordinates are widely used.

In Bilyalov's formalism, on the other hand,
this pair is $(\psi(x), \tensor{P}{^\mu_\nu}(x))$,
where $\tensor{P}{^\mu_\nu}(x)$
is a uniquely-specified general coordinate transformation that maps
the metric $g_{\mu\nu}(x)$ to $\eta_{\mu\nu}$.
One could think of
$\tensor{P}{^\mu_a}(x)$
as a preferred vierbein, since the vierbein does exactly
what $P$ is constructed to, via Eq.~(\ref{eqn:vb:metric}).
Thus, in Bilyalov's formalism,
the spinor field effectively transforms according to a change of tetrad basis,
with the change of basis determined by the preferred vierbein
in each coordinate system.
(The formalism can also accommodate transformation to light front coordinates,
which requires the coordinate swap matrix $T$ that Bilyalov defines in
Ref.~\cite{Bilyalov:1992fd}---which is precisely what's missing
in the Ogievetsky-Polubarinov formalism.)
However, spinors transform in Bilyalov's formalism \emph{only} under
Lorentz transformations---and thus remain invariant under dilations,
giving them the wrong transformation properties under the
special conformal group.
(The standard tetrad formalism and the Ogievetsky-Polubarinov framework
both correctly account for scaling under dilations, by contrast.
See Pitts~\cite{Pitts:2011jv} for more information about the latter.)

Perhaps within these non-standard frameworks,
the derivation of Ref.~\cite{Freese:2021jqs} may be valid after all;
it is merely wrong only the standard formalism.
At the same time, the alternate frameworks
may require refinement before they're fully applicable.


\bibliography{references.bib}

\begin{thebibliography}{90}%
\makeatletter
\providecommand \@ifxundefined [1]{%
 \@ifx{#1\undefined}
}%
\providecommand \@ifnum [1]{%
 \ifnum #1\expandafter \@firstoftwo
 \else \expandafter \@secondoftwo
 \fi
}%
\providecommand \@ifx [1]{%
 \ifx #1\expandafter \@firstoftwo
 \else \expandafter \@secondoftwo
 \fi
}%
\providecommand \natexlab [1]{#1}%
\providecommand \enquote  [1]{``#1''}%
\providecommand \bibnamefont  [1]{#1}%
\providecommand \bibfnamefont [1]{#1}%
\providecommand \citenamefont [1]{#1}%
\providecommand \href@noop [0]{\@secondoftwo}%
\providecommand \href [0]{\begingroup \@sanitize@url \@href}%
\providecommand \@href[1]{\@@startlink{#1}\@@href}%
\providecommand \@@href[1]{\endgroup#1\@@endlink}%
\providecommand \@sanitize@url [0]{\catcode `\\12\catcode `\$12\catcode
  `\&12\catcode `\#12\catcode `\^12\catcode `\_12\catcode `\%12\relax}%
\providecommand \@@startlink[1]{}%
\providecommand \@@endlink[0]{}%
\providecommand \url  [0]{\begingroup\@sanitize@url \@url }%
\providecommand \@url [1]{\endgroup\@href {#1}{\urlprefix }}%
\providecommand \urlprefix  [0]{URL }%
\providecommand \Eprint [0]{\href }%
\providecommand \doibase [0]{http://dx.doi.org/}%
\providecommand \selectlanguage [0]{\@gobble}%
\providecommand \bibinfo  [0]{\@secondoftwo}%
\providecommand \bibfield  [0]{\@secondoftwo}%
\providecommand \translation [1]{[#1]}%
\providecommand \BibitemOpen [0]{}%
\providecommand \bibitemStop [0]{}%
\providecommand \bibitemNoStop [0]{.\EOS\space}%
\providecommand \EOS [0]{\spacefactor3000\relax}%
\providecommand \BibitemShut  [1]{\csname bibitem#1\endcsname}%
\let\auto@bib@innerbib\@empty
\bibitem [{\citenamefont {Freese}(2022)}]{Freese:2021jqs}%
  \BibitemOpen
  \bibfield  {author} {\bibinfo {author} {\bibfnamefont {Adam}\ \bibnamefont
  {Freese}},\ }\bibfield  {title} {\enquote {\bibinfo {title}
  {{Noether\textquoteright{}s theorems and the energy-momentum tensor in
  quantum gauge theories}},}\ }\href {\doibase 10.1103/PhysRevD.106.125012}
  {\bibfield  {journal} {\bibinfo  {journal} {Phys. Rev. D}\ }\textbf {\bibinfo
  {volume} {106}},\ \bibinfo {pages} {125012} (\bibinfo {year} {2022})},\
  \Eprint {http://arxiv.org/abs/2112.00047} {arXiv:2112.00047 [hep-th]}
  \BibitemShut {NoStop}%
\bibitem [{\citenamefont {Ji}(1995)}]{Ji:1995sv}%
  \BibitemOpen
  \bibfield  {author} {\bibinfo {author} {\bibfnamefont {Xiang-Dong}\
  \bibnamefont {Ji}},\ }\bibfield  {title} {\enquote {\bibinfo {title}
  {{Breakup of hadron masses and energy - momentum tensor of QCD}},}\ }\href
  {\doibase 10.1103/PhysRevD.52.271} {\bibfield  {journal} {\bibinfo  {journal}
  {Phys. Rev. D}\ }\textbf {\bibinfo {volume} {52}},\ \bibinfo {pages}
  {271--281} (\bibinfo {year} {1995})},\ \Eprint
  {http://arxiv.org/abs/hep-ph/9502213} {arXiv:hep-ph/9502213} \BibitemShut
  {NoStop}%
\bibitem [{\citenamefont {Lorc\'e}(2018)}]{Lorce:2017xzd}%
  \BibitemOpen
  \bibfield  {author} {\bibinfo {author} {\bibfnamefont {C\'edric}\
  \bibnamefont {Lorc\'e}},\ }\bibfield  {title} {\enquote {\bibinfo {title}
  {{On the hadron mass decomposition}},}\ }\href {\doibase
  10.1140/epjc/s10052-018-5561-2} {\bibfield  {journal} {\bibinfo  {journal}
  {Eur. Phys. J. C}\ }\textbf {\bibinfo {volume} {78}},\ \bibinfo {pages} {120}
  (\bibinfo {year} {2018})},\ \Eprint {http://arxiv.org/abs/1706.05853}
  {arXiv:1706.05853 [hep-ph]} \BibitemShut {NoStop}%
\bibitem [{\citenamefont {Metz}\ \emph {et~al.}(2020)\citenamefont {Metz},
  \citenamefont {Pasquini},\ and\ \citenamefont {Rodini}}]{Metz:2020vxd}%
  \BibitemOpen
  \bibfield  {author} {\bibinfo {author} {\bibfnamefont {Andreas}\ \bibnamefont
  {Metz}}, \bibinfo {author} {\bibfnamefont {Barbara}\ \bibnamefont
  {Pasquini}}, \ and\ \bibinfo {author} {\bibfnamefont {Simone}\ \bibnamefont
  {Rodini}},\ }\bibfield  {title} {\enquote {\bibinfo {title} {{Revisiting the
  proton mass decomposition}},}\ }\href {\doibase 10.1103/PhysRevD.102.114042}
  {\bibfield  {journal} {\bibinfo  {journal} {Phys. Rev. D}\ }\textbf {\bibinfo
  {volume} {102}},\ \bibinfo {pages} {114042} (\bibinfo {year} {2020})},\
  \Eprint {http://arxiv.org/abs/2006.11171} {arXiv:2006.11171 [hep-ph]}
  \BibitemShut {NoStop}%
\bibitem [{\citenamefont {Ji}(2021)}]{Ji:2021mtz}%
  \BibitemOpen
  \bibfield  {author} {\bibinfo {author} {\bibfnamefont {Xiangdong}\
  \bibnamefont {Ji}},\ }\bibfield  {title} {\enquote {\bibinfo {title} {{Proton
  mass decomposition: naturalness and interpretations}},}\ }\href {\doibase
  10.1007/s11467-021-1065-x} {\bibfield  {journal} {\bibinfo  {journal} {Front.
  Phys. (Beijing)}\ }\textbf {\bibinfo {volume} {16}},\ \bibinfo {pages}
  {64601} (\bibinfo {year} {2021})},\ \Eprint {http://arxiv.org/abs/2102.07830}
  {arXiv:2102.07830 [hep-ph]} \BibitemShut {NoStop}%
\bibitem [{\citenamefont {Lorc\'e}\ \emph {et~al.}(2021)\citenamefont
  {Lorc\'e}, \citenamefont {Metz}, \citenamefont {Pasquini},\ and\
  \citenamefont {Rodini}}]{Lorce:2021xku}%
  \BibitemOpen
  \bibfield  {author} {\bibinfo {author} {\bibfnamefont {C\'edric}\
  \bibnamefont {Lorc\'e}}, \bibinfo {author} {\bibfnamefont {Andreas}\
  \bibnamefont {Metz}}, \bibinfo {author} {\bibfnamefont {Barbara}\
  \bibnamefont {Pasquini}}, \ and\ \bibinfo {author} {\bibfnamefont {Simone}\
  \bibnamefont {Rodini}},\ }\bibfield  {title} {\enquote {\bibinfo {title}
  {{Energy-momentum tensor in QCD: nucleon mass decomposition and mechanical
  equilibrium}},}\ }\href {\doibase 10.1007/JHEP11(2021)121} {\bibfield
  {journal} {\bibinfo  {journal} {JHEP}\ }\textbf {\bibinfo {volume} {11}},\
  \bibinfo {pages} {121} (\bibinfo {year} {2021})},\ \Eprint
  {http://arxiv.org/abs/2109.11785} {arXiv:2109.11785 [hep-ph]} \BibitemShut
  {NoStop}%
\bibitem [{\citenamefont {Jaffe}\ and\ \citenamefont
  {Manohar}(1990)}]{Jaffe:1989jz}%
  \BibitemOpen
  \bibfield  {author} {\bibinfo {author} {\bibfnamefont {R.~L.}\ \bibnamefont
  {Jaffe}}\ and\ \bibinfo {author} {\bibfnamefont {Aneesh}\ \bibnamefont
  {Manohar}},\ }\bibfield  {title} {\enquote {\bibinfo {title} {{The $g_1$
  Problem: Fact and Fantasy on the Spin of the Proton}},}\ }\href {\doibase
  10.1016/0550-3213(90)90506-9} {\bibfield  {journal} {\bibinfo  {journal}
  {Nucl. Phys. B}\ }\textbf {\bibinfo {volume} {337}},\ \bibinfo {pages}
  {509--546} (\bibinfo {year} {1990})}\BibitemShut {NoStop}%
\bibitem [{\citenamefont {Ji}(1997)}]{Ji:1996ek}%
  \BibitemOpen
  \bibfield  {author} {\bibinfo {author} {\bibfnamefont {Xiang-Dong}\
  \bibnamefont {Ji}},\ }\bibfield  {title} {\enquote {\bibinfo {title}
  {{Gauge-Invariant Decomposition of Nucleon Spin}},}\ }\href {\doibase
  10.1103/PhysRevLett.78.610} {\bibfield  {journal} {\bibinfo  {journal} {Phys.
  Rev. Lett.}\ }\textbf {\bibinfo {volume} {78}},\ \bibinfo {pages} {610--613}
  (\bibinfo {year} {1997})},\ \Eprint {http://arxiv.org/abs/hep-ph/9603249}
  {arXiv:hep-ph/9603249} \BibitemShut {NoStop}%
\bibitem [{\citenamefont {Leader}\ and\ \citenamefont
  {Lorc\'e}(2014)}]{Leader:2013jra}%
  \BibitemOpen
  \bibfield  {author} {\bibinfo {author} {\bibfnamefont {E.}~\bibnamefont
  {Leader}}\ and\ \bibinfo {author} {\bibfnamefont {C.}~\bibnamefont
  {Lorc\'e}},\ }\bibfield  {title} {\enquote {\bibinfo {title} {{The angular
  momentum controversy: What\textquoteright{}s it all about and does it
  matter?}}}\ }\href {\doibase 10.1016/j.physrep.2014.02.010} {\bibfield
  {journal} {\bibinfo  {journal} {Phys. Rept.}\ }\textbf {\bibinfo {volume}
  {541}},\ \bibinfo {pages} {163--248} (\bibinfo {year} {2014})},\ \Eprint
  {http://arxiv.org/abs/1309.4235} {arXiv:1309.4235 [hep-ph]} \BibitemShut
  {NoStop}%
\bibitem [{\citenamefont {Polyakov}(2003)}]{Polyakov:2002yz}%
  \BibitemOpen
  \bibfield  {author} {\bibinfo {author} {\bibfnamefont {M.~V.}\ \bibnamefont
  {Polyakov}},\ }\bibfield  {title} {\enquote {\bibinfo {title} {{Generalized
  parton distributions and strong forces inside nucleons and nuclei}},}\ }\href
  {\doibase 10.1016/S0370-2693(03)00036-4} {\bibfield  {journal} {\bibinfo
  {journal} {Phys. Lett. B}\ }\textbf {\bibinfo {volume} {555}},\ \bibinfo
  {pages} {57--62} (\bibinfo {year} {2003})},\ \Eprint
  {http://arxiv.org/abs/hep-ph/0210165} {arXiv:hep-ph/0210165} \BibitemShut
  {NoStop}%
\bibitem [{\citenamefont {Polyakov}\ and\ \citenamefont
  {Schweitzer}(2018)}]{Polyakov:2018zvc}%
  \BibitemOpen
  \bibfield  {author} {\bibinfo {author} {\bibfnamefont {Maxim~V.}\
  \bibnamefont {Polyakov}}\ and\ \bibinfo {author} {\bibfnamefont {Peter}\
  \bibnamefont {Schweitzer}},\ }\bibfield  {title} {\enquote {\bibinfo {title}
  {{Forces inside hadrons: pressure, surface tension, mechanical radius, and
  all that}},}\ }\href {\doibase 10.1142/S0217751X18300259} {\bibfield
  {journal} {\bibinfo  {journal} {Int. J. Mod. Phys. A}\ }\textbf {\bibinfo
  {volume} {33}},\ \bibinfo {pages} {1830025} (\bibinfo {year} {2018})},\
  \Eprint {http://arxiv.org/abs/1805.06596} {arXiv:1805.06596 [hep-ph]}
  \BibitemShut {NoStop}%
\bibitem [{\citenamefont {Lorc\'e}\ \emph {et~al.}(2019)\citenamefont
  {Lorc\'e}, \citenamefont {Moutarde},\ and\ \citenamefont
  {Trawi\'nski}}]{Lorce:2018egm}%
  \BibitemOpen
  \bibfield  {author} {\bibinfo {author} {\bibfnamefont {C\'edric}\
  \bibnamefont {Lorc\'e}}, \bibinfo {author} {\bibfnamefont {Herv\'e}\
  \bibnamefont {Moutarde}}, \ and\ \bibinfo {author} {\bibfnamefont
  {Arkadiusz~P.}\ \bibnamefont {Trawi\'nski}},\ }\bibfield  {title} {\enquote
  {\bibinfo {title} {{Revisiting the mechanical properties of the nucleon}},}\
  }\href {\doibase 10.1140/epjc/s10052-019-6572-3} {\bibfield  {journal}
  {\bibinfo  {journal} {Eur. Phys. J. C}\ }\textbf {\bibinfo {volume} {79}},\
  \bibinfo {pages} {89} (\bibinfo {year} {2019})},\ \Eprint
  {http://arxiv.org/abs/1810.09837} {arXiv:1810.09837 [hep-ph]} \BibitemShut
  {NoStop}%
\bibitem [{\citenamefont {Ji}\ and\ \citenamefont {Liu}(2022)}]{Ji:2021mfb}%
  \BibitemOpen
  \bibfield  {author} {\bibinfo {author} {\bibfnamefont {Xiangdong}\
  \bibnamefont {Ji}}\ and\ \bibinfo {author} {\bibfnamefont {Yizhuang}\
  \bibnamefont {Liu}},\ }\bibfield  {title} {\enquote {\bibinfo {title}
  {{Momentum-Current Gravitational Multipoles of Hadrons}},}\ }\href {\doibase
  10.1103/PhysRevD.106.034028} {\bibfield  {journal} {\bibinfo  {journal}
  {Phys. Rev. D}\ }\textbf {\bibinfo {volume} {106}},\ \bibinfo {pages}
  {034028} (\bibinfo {year} {2022})},\ \Eprint
  {http://arxiv.org/abs/2110.14781} {arXiv:2110.14781 [hep-ph]} \BibitemShut
  {NoStop}%
\bibitem [{\citenamefont {Ji}\ \emph {et~al.}(2024)\citenamefont {Ji},
  \citenamefont {Yang},\ and\ \citenamefont {Liu}}]{Ji:2022exr}%
  \BibitemOpen
  \bibfield  {author} {\bibinfo {author} {\bibfnamefont {Xiangdong}\
  \bibnamefont {Ji}}, \bibinfo {author} {\bibfnamefont {Jinghong}\ \bibnamefont
  {Yang}}, \ and\ \bibinfo {author} {\bibfnamefont {Yizhuang}\ \bibnamefont
  {Liu}},\ }\bibfield  {title} {\enquote {\bibinfo {title} {{Gravitational
  tensor-monopole moment of the hydrogen atom to order
  O(\ensuremath{\alpha})}},}\ }\href {\doibase 10.1103/PhysRevD.110.114045}
  {\bibfield  {journal} {\bibinfo  {journal} {Phys. Rev. D}\ }\textbf {\bibinfo
  {volume} {110}},\ \bibinfo {pages} {114045} (\bibinfo {year} {2024})},\
  \Eprint {http://arxiv.org/abs/2208.05029} {arXiv:2208.05029 [hep-ph]}
  \BibitemShut {NoStop}%
\bibitem [{\citenamefont {Ji}\ and\ \citenamefont {Yang}(2025)}]{Ji:2025gsq}%
  \BibitemOpen
  \bibfield  {author} {\bibinfo {author} {\bibfnamefont {Xiangdong}\
  \bibnamefont {Ji}}\ and\ \bibinfo {author} {\bibfnamefont {Chen}\
  \bibnamefont {Yang}},\ }\bibfield  {title} {\enquote {\bibinfo {title}
  {{Momentum Flow and Forces on Quarks in the Nucleon}},}\ }\href@noop {} {\
  (\bibinfo {year} {2025})},\ \Eprint {http://arxiv.org/abs/2503.01991}
  {arXiv:2503.01991 [hep-ph]} \BibitemShut {NoStop}%
\bibitem [{\citenamefont {Burkert}\ \emph {et~al.}(2023)\citenamefont
  {Burkert}, \citenamefont {Elouadrhiri}, \citenamefont {Girod}, \citenamefont
  {Lorc\'e}, \citenamefont {Schweitzer},\ and\ \citenamefont
  {Shanahan}}]{Burkert:2023wzr}%
  \BibitemOpen
  \bibfield  {author} {\bibinfo {author} {\bibfnamefont {V.~D.}\ \bibnamefont
  {Burkert}}, \bibinfo {author} {\bibfnamefont {L.}~\bibnamefont
  {Elouadrhiri}}, \bibinfo {author} {\bibfnamefont {F.~X.}\ \bibnamefont
  {Girod}}, \bibinfo {author} {\bibfnamefont {C.}~\bibnamefont {Lorc\'e}},
  \bibinfo {author} {\bibfnamefont {P.}~\bibnamefont {Schweitzer}}, \ and\
  \bibinfo {author} {\bibfnamefont {P.~E.}\ \bibnamefont {Shanahan}},\
  }\bibfield  {title} {\enquote {\bibinfo {title} {{Colloquium: Gravitational
  form factors of the proton}},}\ }\href {\doibase
  10.1103/RevModPhys.95.041002} {\bibfield  {journal} {\bibinfo  {journal}
  {Rev. Mod. Phys.}\ }\textbf {\bibinfo {volume} {95}},\ \bibinfo {pages}
  {041002} (\bibinfo {year} {2023})},\ \Eprint
  {http://arxiv.org/abs/2303.08347} {arXiv:2303.08347 [hep-ph]} \BibitemShut
  {NoStop}%
\bibitem [{\citenamefont {Freese}(2025{\natexlab{a}})}]{Freese:2024rkr}%
  \BibitemOpen
  \bibfield  {author} {\bibinfo {author} {\bibfnamefont {Adam}\ \bibnamefont
  {Freese}},\ }\bibfield  {title} {\enquote {\bibinfo {title} {{Quantum
  stresses in the hydrogen atom}},}\ }\href {\doibase
  10.1103/PhysRevD.111.034047} {\bibfield  {journal} {\bibinfo  {journal}
  {Phys. Rev. D}\ }\textbf {\bibinfo {volume} {111}},\ \bibinfo {pages}
  {034047} (\bibinfo {year} {2025}{\natexlab{a}})},\ \Eprint
  {http://arxiv.org/abs/2412.09664} {arXiv:2412.09664 [hep-ph]} \BibitemShut
  {NoStop}%
\bibitem [{\citenamefont {Lorc\'e}\ and\ \citenamefont
  {Schweitzer}(2025)}]{Lorce:2025oot}%
  \BibitemOpen
  \bibfield  {author} {\bibinfo {author} {\bibfnamefont {C\'edric}\
  \bibnamefont {Lorc\'e}}\ and\ \bibinfo {author} {\bibfnamefont {Peter}\
  \bibnamefont {Schweitzer}},\ }\bibfield  {title} {\enquote {\bibinfo {title}
  {{Pressure inside hadrons: criticism, conjectures, and all that}},}\ }\href
  {\doibase 10.5506/APhysPolB.56.3-A17} {\bibfield  {journal} {\bibinfo
  {journal} {Acta Phys. Polon. B}\ }\textbf {\bibinfo {volume} {56}},\ \bibinfo
  {pages} {3--A17} (\bibinfo {year} {2025})},\ \Eprint
  {http://arxiv.org/abs/2501.04622} {arXiv:2501.04622 [hep-ph]} \BibitemShut
  {NoStop}%
\bibitem [{\citenamefont {Burkert}\ \emph {et~al.}(2018)\citenamefont
  {Burkert}, \citenamefont {Elouadrhiri},\ and\ \citenamefont
  {Girod}}]{Burkert:2018bqq}%
  \BibitemOpen
  \bibfield  {author} {\bibinfo {author} {\bibfnamefont {V.~D.}\ \bibnamefont
  {Burkert}}, \bibinfo {author} {\bibfnamefont {L.}~\bibnamefont
  {Elouadrhiri}}, \ and\ \bibinfo {author} {\bibfnamefont {F.~X.}\ \bibnamefont
  {Girod}},\ }\bibfield  {title} {\enquote {\bibinfo {title} {{The pressure
  distribution inside the proton}},}\ }\href {\doibase
  10.1038/s41586-018-0060-z} {\bibfield  {journal} {\bibinfo  {journal}
  {Nature}\ }\textbf {\bibinfo {volume} {557}},\ \bibinfo {pages} {396--399}
  (\bibinfo {year} {2018})}\BibitemShut {NoStop}%
\bibitem [{\citenamefont {Kumeri\v{c}ki}(2019)}]{Kumericki:2019ddg}%
  \BibitemOpen
  \bibfield  {author} {\bibinfo {author} {\bibfnamefont {Kre\v{s}imir}\
  \bibnamefont {Kumeri\v{c}ki}},\ }\bibfield  {title} {\enquote {\bibinfo
  {title} {{Measurability of pressure inside the proton}},}\ }\href {\doibase
  10.1038/s41586-019-1211-6} {\bibfield  {journal} {\bibinfo  {journal}
  {Nature}\ }\textbf {\bibinfo {volume} {570}},\ \bibinfo {pages} {E1--E2}
  (\bibinfo {year} {2019})}\BibitemShut {NoStop}%
\bibitem [{\citenamefont {Burkert}\ \emph {et~al.}(2021)\citenamefont
  {Burkert}, \citenamefont {Elouadrhiri},\ and\ \citenamefont
  {Girod}}]{Burkert:2021ith}%
  \BibitemOpen
  \bibfield  {author} {\bibinfo {author} {\bibfnamefont {V.~D.}\ \bibnamefont
  {Burkert}}, \bibinfo {author} {\bibfnamefont {L.}~\bibnamefont
  {Elouadrhiri}}, \ and\ \bibinfo {author} {\bibfnamefont {F.~X.}\ \bibnamefont
  {Girod}},\ }\bibfield  {title} {\enquote {\bibinfo {title} {{Determination of
  shear forces inside the proton}},}\ }\href@noop {} {\  (\bibinfo {year}
  {2021})},\ \Eprint {http://arxiv.org/abs/2104.02031} {arXiv:2104.02031
  [nucl-ex]} \BibitemShut {NoStop}%
\bibitem [{\citenamefont {Duran}\ \emph {et~al.}(2023)\citenamefont {Duran}
  \emph {et~al.}}]{Duran:2022xag}%
  \BibitemOpen
  \bibfield  {author} {\bibinfo {author} {\bibfnamefont {B.}~\bibnamefont
  {Duran}} \emph {et~al.},\ }\bibfield  {title} {\enquote {\bibinfo {title}
  {{Determining the gluonic gravitational form factors of the proton}},}\
  }\href {\doibase 10.1038/s41586-023-05730-4} {\bibfield  {journal} {\bibinfo
  {journal} {Nature}\ }\textbf {\bibinfo {volume} {615}},\ \bibinfo {pages}
  {813--816} (\bibinfo {year} {2023})},\ \Eprint
  {http://arxiv.org/abs/2207.05212} {arXiv:2207.05212 [nucl-ex]} \BibitemShut
  {NoStop}%
\bibitem [{\citenamefont {Guo}\ \emph {et~al.}(2025)\citenamefont {Guo},
  \citenamefont {Yuan},\ and\ \citenamefont {Zhao}}]{Guo:2025jiz}%
  \BibitemOpen
  \bibfield  {author} {\bibinfo {author} {\bibfnamefont {Yuxun}\ \bibnamefont
  {Guo}}, \bibinfo {author} {\bibfnamefont {Feng}\ \bibnamefont {Yuan}}, \ and\
  \bibinfo {author} {\bibfnamefont {Wenbin}\ \bibnamefont {Zhao}},\ }\bibfield
  {title} {\enquote {\bibinfo {title} {{Bayesian Inferring Nucleon's
  Gravitation Form Factors via Near-threshold $J/\psi$ Photoproduction}},}\
  }\href@noop {} {\  (\bibinfo {year} {2025})},\ \Eprint
  {http://arxiv.org/abs/2501.10532} {arXiv:2501.10532 [hep-ph]} \BibitemShut
  {NoStop}%
\bibitem [{\citenamefont {Hatta}\ \emph {et~al.}(2025)\citenamefont {Hatta},
  \citenamefont {Klest}, \citenamefont {Passek-K.},\ and\ \citenamefont
  {Schoenleber}}]{Hatta:2025vhs}%
  \BibitemOpen
  \bibfield  {author} {\bibinfo {author} {\bibfnamefont {Yoshitaka}\
  \bibnamefont {Hatta}}, \bibinfo {author} {\bibfnamefont {Henry~T.}\
  \bibnamefont {Klest}}, \bibinfo {author} {\bibfnamefont {Kornelija}\
  \bibnamefont {Passek-K.}}, \ and\ \bibinfo {author} {\bibfnamefont {Jakob}\
  \bibnamefont {Schoenleber}},\ }\bibfield  {title} {\enquote {\bibinfo {title}
  {{Deeply virtual $\phi$-meson production near threshold}},}\ }\href@noop {}
  {\  (\bibinfo {year} {2025})},\ \Eprint {http://arxiv.org/abs/2501.12343}
  {arXiv:2501.12343 [hep-ph]} \BibitemShut {NoStop}%
\bibitem [{\citenamefont {Shanahan}\ and\ \citenamefont
  {Detmold}(2019{\natexlab{a}})}]{Shanahan:2018pib}%
  \BibitemOpen
  \bibfield  {author} {\bibinfo {author} {\bibfnamefont {P.~E.}\ \bibnamefont
  {Shanahan}}\ and\ \bibinfo {author} {\bibfnamefont {W.}~\bibnamefont
  {Detmold}},\ }\bibfield  {title} {\enquote {\bibinfo {title} {{Gluon
  gravitational form factors of the nucleon and the pion from lattice QCD}},}\
  }\href {\doibase 10.1103/PhysRevD.99.014511} {\bibfield  {journal} {\bibinfo
  {journal} {Phys. Rev. D}\ }\textbf {\bibinfo {volume} {99}},\ \bibinfo
  {pages} {014511} (\bibinfo {year} {2019}{\natexlab{a}})},\ \Eprint
  {http://arxiv.org/abs/1810.04626} {arXiv:1810.04626 [hep-lat]} \BibitemShut
  {NoStop}%
\bibitem [{\citenamefont {Shanahan}\ and\ \citenamefont
  {Detmold}(2019{\natexlab{b}})}]{Shanahan:2018nnv}%
  \BibitemOpen
  \bibfield  {author} {\bibinfo {author} {\bibfnamefont {P.~E.}\ \bibnamefont
  {Shanahan}}\ and\ \bibinfo {author} {\bibfnamefont {W.}~\bibnamefont
  {Detmold}},\ }\bibfield  {title} {\enquote {\bibinfo {title} {{Pressure
  Distribution and Shear Forces inside the Proton}},}\ }\href {\doibase
  10.1103/PhysRevLett.122.072003} {\bibfield  {journal} {\bibinfo  {journal}
  {Phys. Rev. Lett.}\ }\textbf {\bibinfo {volume} {122}},\ \bibinfo {pages}
  {072003} (\bibinfo {year} {2019}{\natexlab{b}})},\ \Eprint
  {http://arxiv.org/abs/1810.07589} {arXiv:1810.07589 [nucl-th]} \BibitemShut
  {NoStop}%
\bibitem [{\citenamefont {Pefkou}\ \emph {et~al.}(2022)\citenamefont {Pefkou},
  \citenamefont {Hackett},\ and\ \citenamefont {Shanahan}}]{Pefkou:2021fni}%
  \BibitemOpen
  \bibfield  {author} {\bibinfo {author} {\bibfnamefont {Dimitra~A.}\
  \bibnamefont {Pefkou}}, \bibinfo {author} {\bibfnamefont {Daniel~C.}\
  \bibnamefont {Hackett}}, \ and\ \bibinfo {author} {\bibfnamefont {Phiala~E.}\
  \bibnamefont {Shanahan}},\ }\bibfield  {title} {\enquote {\bibinfo {title}
  {{Gluon gravitational structure of hadrons of different spin}},}\ }\href
  {\doibase 10.1103/PhysRevD.105.054509} {\bibfield  {journal} {\bibinfo
  {journal} {Phys. Rev. D}\ }\textbf {\bibinfo {volume} {105}},\ \bibinfo
  {pages} {054509} (\bibinfo {year} {2022})},\ \Eprint
  {http://arxiv.org/abs/2107.10368} {arXiv:2107.10368 [hep-lat]} \BibitemShut
  {NoStop}%
\bibitem [{\citenamefont {Hackett}\ \emph {et~al.}(2024)\citenamefont
  {Hackett}, \citenamefont {Pefkou},\ and\ \citenamefont
  {Shanahan}}]{Hackett:2023rif}%
  \BibitemOpen
  \bibfield  {author} {\bibinfo {author} {\bibfnamefont {Daniel~C.}\
  \bibnamefont {Hackett}}, \bibinfo {author} {\bibfnamefont {Dimitra~A.}\
  \bibnamefont {Pefkou}}, \ and\ \bibinfo {author} {\bibfnamefont {Phiala~E.}\
  \bibnamefont {Shanahan}},\ }\bibfield  {title} {\enquote {\bibinfo {title}
  {{Gravitational Form Factors of the Proton from Lattice QCD}},}\ }\href
  {\doibase 10.1103/PhysRevLett.132.251904} {\bibfield  {journal} {\bibinfo
  {journal} {Phys. Rev. Lett.}\ }\textbf {\bibinfo {volume} {132}},\ \bibinfo
  {pages} {251904} (\bibinfo {year} {2024})},\ \Eprint
  {http://arxiv.org/abs/2310.08484} {arXiv:2310.08484 [hep-lat]} \BibitemShut
  {NoStop}%
\bibitem [{\citenamefont {Pefkou}(2023)}]{Pefkou:2023okb}%
  \BibitemOpen
  \bibfield  {author} {\bibinfo {author} {\bibfnamefont {Dimitra~Anastasia}\
  \bibnamefont {Pefkou}},\ }\emph {\bibinfo {title} {{Gravitational form
  factors of hadrons from lattice QCD}}},\ \href@noop {} {Ph.D. thesis},\
  \bibinfo  {school} {MIT} (\bibinfo {year} {2023})\BibitemShut {NoStop}%
\bibitem [{\citenamefont {Hackett}\ \emph {et~al.}(2023)\citenamefont
  {Hackett}, \citenamefont {Oare}, \citenamefont {Pefkou},\ and\ \citenamefont
  {Shanahan}}]{Hackett:2023nkr}%
  \BibitemOpen
  \bibfield  {author} {\bibinfo {author} {\bibfnamefont {Daniel~C.}\
  \bibnamefont {Hackett}}, \bibinfo {author} {\bibfnamefont {Patrick~R.}\
  \bibnamefont {Oare}}, \bibinfo {author} {\bibfnamefont {Dimitra~A.}\
  \bibnamefont {Pefkou}}, \ and\ \bibinfo {author} {\bibfnamefont {Phiala~E.}\
  \bibnamefont {Shanahan}},\ }\bibfield  {title} {\enquote {\bibinfo {title}
  {{Gravitational form factors of the pion from lattice QCD}},}\ }\href
  {\doibase 10.1103/PhysRevD.108.114504} {\bibfield  {journal} {\bibinfo
  {journal} {Phys. Rev. D}\ }\textbf {\bibinfo {volume} {108}},\ \bibinfo
  {pages} {114504} (\bibinfo {year} {2023})},\ \Eprint
  {http://arxiv.org/abs/2307.11707} {arXiv:2307.11707 [hep-lat]} \BibitemShut
  {NoStop}%
\bibitem [{\citenamefont {{Belinfante}}(1939)}]{Belinfante:1939emt}%
  \BibitemOpen
  \bibfield  {author} {\bibinfo {author} {\bibfnamefont {F.~J.}\ \bibnamefont
  {{Belinfante}}},\ }\bibfield  {title} {\enquote {\bibinfo {title} {{On the
  spin angular momentum of mesons}},}\ }\href {\doibase
  10.1016/S0031-8914(39)90090-X} {\bibfield  {journal} {\bibinfo  {journal}
  {Physica}\ }\textbf {\bibinfo {volume} {6}},\ \bibinfo {pages} {887--898}
  (\bibinfo {year} {1939})}\BibitemShut {NoStop}%
\bibitem [{\citenamefont {Lorc{\'e}}\ \emph {et~al.}(2025)\citenamefont
  {Lorc{\'e}}, \citenamefont {Mukherjee}, \citenamefont {Singh},\ and\
  \citenamefont {Won}}]{Lorce:2025pxt}%
  \BibitemOpen
  \bibfield  {author} {\bibinfo {author} {\bibfnamefont {C{\'e}dric}\
  \bibnamefont {Lorc{\'e}}}, \bibinfo {author} {\bibfnamefont {Asmita}\
  \bibnamefont {Mukherjee}}, \bibinfo {author} {\bibfnamefont {Ravi}\
  \bibnamefont {Singh}}, \ and\ \bibinfo {author} {\bibfnamefont {Ho-Yeon}\
  \bibnamefont {Won}},\ }\bibfield  {title} {\enquote {\bibinfo {title}
  {{Mapping the transverse spin sum rule in position space}},}\ }\href
  {\doibase 10.1016/j.physletb.2025.139792} {\bibfield  {journal} {\bibinfo
  {journal} {Phys. Lett. B}\ }\textbf {\bibinfo {volume} {868}},\ \bibinfo
  {pages} {139792} (\bibinfo {year} {2025})},\ \Eprint
  {http://arxiv.org/abs/2505.20468} {arXiv:2505.20468 [hep-ph]} \BibitemShut
  {NoStop}%
\bibitem [{\citenamefont {Won}\ and\ \citenamefont
  {Lorc{\'e}}(2025)}]{Won:2025dgc}%
  \BibitemOpen
  \bibfield  {author} {\bibinfo {author} {\bibfnamefont {Ho-Yeon}\ \bibnamefont
  {Won}}\ and\ \bibinfo {author} {\bibfnamefont {C{\'e}dric}\ \bibnamefont
  {Lorc{\'e}}},\ }\bibfield  {title} {\enquote {\bibinfo {title} {{Relativistic
  energy-momentum tensor distributions in a polarized nucleon}},}\ }\href
  {\doibase 10.1103/PhysRevD.111.094021} {\bibfield  {journal} {\bibinfo
  {journal} {Phys. Rev. D}\ }\textbf {\bibinfo {volume} {111}},\ \bibinfo
  {pages} {094021} (\bibinfo {year} {2025})},\ \Eprint
  {http://arxiv.org/abs/2503.07382} {arXiv:2503.07382 [hep-ph]} \BibitemShut
  {NoStop}%
\bibitem [{\citenamefont {Freese}\ and\ \citenamefont
  {Miller}(2023)}]{Freese:2023abr}%
  \BibitemOpen
  \bibfield  {author} {\bibinfo {author} {\bibfnamefont {Adam}\ \bibnamefont
  {Freese}}\ and\ \bibinfo {author} {\bibfnamefont {Gerald~A.}\ \bibnamefont
  {Miller}},\ }\bibfield  {title} {\enquote {\bibinfo {title} {{Synchronization
  effects on rest frame energy and momentum densities in the proton}},}\ }\href
  {\doibase 10.1103/PhysRevD.108.094026} {\bibfield  {journal} {\bibinfo
  {journal} {Phys. Rev. D}\ }\textbf {\bibinfo {volume} {108}},\ \bibinfo
  {pages} {094026} (\bibinfo {year} {2023})},\ \Eprint
  {http://arxiv.org/abs/2307.11165} {arXiv:2307.11165 [hep-ph]} \BibitemShut
  {NoStop}%
\bibitem [{\citenamefont {Freese}(2025{\natexlab{b}})}]{Freese:2025tqd}%
  \BibitemOpen
  \bibfield  {author} {\bibinfo {author} {\bibfnamefont {Adam}\ \bibnamefont
  {Freese}},\ }\bibfield  {title} {\enquote {\bibinfo {title} {{Mechanical form
  factors and densities of nonrelativistic fermions}},}\ }\href {\doibase
  10.1103/k3hd-3lqv} {\bibfield  {journal} {\bibinfo  {journal} {Phys. Rev. D}\
  }\textbf {\bibinfo {volume} {112}},\ \bibinfo {pages} {034037} (\bibinfo
  {year} {2025}{\natexlab{b}})},\ \Eprint {http://arxiv.org/abs/2505.06135}
  {arXiv:2505.06135 [hep-ph]} \BibitemShut {NoStop}%
\bibitem [{\citenamefont {Kosyakov}(2007)}]{Kosyakov:2007qc}%
  \BibitemOpen
  \bibfield  {author} {\bibinfo {author} {\bibfnamefont {Boris~Pavlovich}\
  \bibnamefont {Kosyakov}},\ }\href
  {https://link.springer.com/book/10.1007/978-3-540-40934-2} {\emph {\bibinfo
  {title} {{Introduction to the Classical Theory of Particles and Fields}}}}\
  (\bibinfo  {publisher} {Springer Berlin Heidelberg New York},\ \bibinfo
  {year} {2007})\BibitemShut {NoStop}%
\bibitem [{\citenamefont {Noether}(1918)}]{Noether:1918zz}%
  \BibitemOpen
  \bibfield  {author} {\bibinfo {author} {\bibfnamefont {Emmy}\ \bibnamefont
  {Noether}},\ }\bibfield  {title} {\enquote {\bibinfo {title} {{Invariant
  Variation Problems}},}\ }\href {\doibase 10.1080/00411457108231446}
  {\bibfield  {journal} {\bibinfo  {journal} {Gott. Nachr.}\ }\textbf {\bibinfo
  {volume} {1918}},\ \bibinfo {pages} {235--257} (\bibinfo {year} {1918})},\
  \Eprint {http://arxiv.org/abs/physics/0503066} {arXiv:physics/0503066}
  \BibitemShut {NoStop}%
\bibitem [{\citenamefont {Weinberg}(1972)}]{Weinberg:1972kfs}%
  \BibitemOpen
  \bibfield  {author} {\bibinfo {author} {\bibfnamefont {Steven}\ \bibnamefont
  {Weinberg}},\ }\href@noop {} {\emph {\bibinfo {title} {{Gravitation and
  Cosmology}: {Principles and Applications of the General Theory of
  Relativity}}}}\ (\bibinfo  {publisher} {John Wiley and Sons},\ \bibinfo
  {address} {New York},\ \bibinfo {year} {1972})\BibitemShut {NoStop}%
\bibitem [{\citenamefont {Carroll}(2019)}]{Carroll:2004st}%
  \BibitemOpen
  \bibfield  {author} {\bibinfo {author} {\bibfnamefont {Sean~M.}\ \bibnamefont
  {Carroll}},\ }\href {\doibase 10.1017/9781108770385} {\emph {\bibinfo {title}
  {{Spacetime and Geometry}: {An Introduction to General Relativity}}}}\
  (\bibinfo  {publisher} {Cambridge University Press},\ \bibinfo {year}
  {2019})\BibitemShut {NoStop}%
\bibitem [{\citenamefont {Hamilton}(2018)}]{hamilton2018general}%
  \BibitemOpen
  \bibfield  {author} {\bibinfo {author} {\bibfnamefont {Andrew~J.S.}\
  \bibnamefont {Hamilton}},\ }\bibfield  {title} {\enquote {\bibinfo {title}
  {General relativity, black holes, and cosmology},}\ }\href
  {https://jila.colorado.edu/~ajsh/astr3740_17/grbook.pdf} {\bibfield
  {journal} {\bibinfo  {journal} {Unpublished}\ } (\bibinfo {year}
  {2018})}\BibitemShut {NoStop}%
\bibitem [{\citenamefont {Bilyalov}(1992)}]{Bilyalov:1992fd}%
  \BibitemOpen
  \bibfield  {author} {\bibinfo {author} {\bibfnamefont {R.~F.}\ \bibnamefont
  {Bilyalov}},\ }\bibfield  {title} {\enquote {\bibinfo {title} {{Conservation
  laws for spinor fields on a Riemannian space-time manifold}},}\ }\href
  {\doibase 10.1007/BF01036530} {\bibfield  {journal} {\bibinfo  {journal}
  {Theor. Math. Phys.}\ }\textbf {\bibinfo {volume} {90}},\ \bibinfo {pages}
  {252--259} (\bibinfo {year} {1992})}\BibitemShut {NoStop}%
\bibitem [{\citenamefont {Bilyalov}(1996)}]{Bilyalov:1996pe}%
  \BibitemOpen
  \bibfield  {author} {\bibinfo {author} {\bibfnamefont {R.~F.}\ \bibnamefont
  {Bilyalov}},\ }\bibfield  {title} {\enquote {\bibinfo {title} {{Symmetric
  energy-momentum tensor of spinor fields}},}\ }\href {\doibase
  10.1007/BF02070677} {\bibfield  {journal} {\bibinfo  {journal} {Theor. Math.
  Phys.}\ }\textbf {\bibinfo {volume} {108}},\ \bibinfo {pages} {1093--1099}
  (\bibinfo {year} {1996})}\BibitemShut {NoStop}%
\bibitem [{\citenamefont {Gamboa~Saravi}(2002)}]{GamboaSaravi:2002vos}%
  \BibitemOpen
  \bibfield  {author} {\bibinfo {author} {\bibfnamefont {Ricardo~E.}\
  \bibnamefont {Gamboa~Saravi}},\ }\bibfield  {title} {\enquote {\bibinfo
  {title} {{The Electromagnetic energy momentum tensor}},}\ }\href {\doibase
  10.1088/0305-4470/35/43/314} {\bibfield  {journal} {\bibinfo  {journal} {J.
  Phys.}\ }\textbf {\bibinfo {volume} {A35}},\ \bibinfo {pages} {9199--9204}
  (\bibinfo {year} {2002})},\ \Eprint {http://arxiv.org/abs/math-ph/0208003}
  {arXiv:math-ph/0208003 [math-ph]} \BibitemShut {NoStop}%
\bibitem [{\citenamefont {Gamboa~Saravi}(2004)}]{GamboaSaravi:2003aq}%
  \BibitemOpen
  \bibfield  {author} {\bibinfo {author} {\bibfnamefont {Ricardo~E.}\
  \bibnamefont {Gamboa~Saravi}},\ }\bibfield  {title} {\enquote {\bibinfo
  {title} {{On the energy momentum tensor}},}\ }\href {\doibase
  10.1088/0305-4470/37/40/017} {\bibfield  {journal} {\bibinfo  {journal} {J.
  Phys.}\ }\textbf {\bibinfo {volume} {A37}},\ \bibinfo {pages} {9573--9586}
  (\bibinfo {year} {2004})},\ \Eprint {http://arxiv.org/abs/math-ph/0306020}
  {arXiv:math-ph/0306020 [math-ph]} \BibitemShut {NoStop}%
\bibitem [{\citenamefont {Zhang}(2005)}]{Zhang:2004jb}%
  \BibitemOpen
  \bibfield  {author} {\bibinfo {author} {\bibfnamefont {Hong-bao}\
  \bibnamefont {Zhang}},\ }\bibfield  {title} {\enquote {\bibinfo {title}
  {{Note on the energy-momentum tensor for general mixed tensor-spinor
  fields}},}\ }\href {\doibase 10.1088/6102/44/6/1007} {\bibfield  {journal}
  {\bibinfo  {journal} {Commun. Theor. Phys.}\ }\textbf {\bibinfo {volume}
  {44}},\ \bibinfo {pages} {1007--1010} (\bibinfo {year} {2005})},\ \Eprint
  {http://arxiv.org/abs/math-ph/0412064} {arXiv:math-ph/0412064} \BibitemShut
  {NoStop}%
\bibitem [{\citenamefont {Helfer}(2016)}]{Helfer:2016zvl}%
  \BibitemOpen
  \bibfield  {author} {\bibinfo {author} {\bibfnamefont {Adam~D.}\ \bibnamefont
  {Helfer}},\ }\bibfield  {title} {\enquote {\bibinfo {title} {{Spinor Lie
  derivatives and Fermion stress\textendash{}energies}},}\ }\href {\doibase
  10.1098/rspa.2015.0757} {\bibfield  {journal} {\bibinfo  {journal} {Proc.
  Roy. Soc. Lond. A}\ }\textbf {\bibinfo {volume} {472}},\ \bibinfo {pages}
  {20150757} (\bibinfo {year} {2016})},\ \Eprint
  {http://arxiv.org/abs/1602.00632} {arXiv:1602.00632 [hep-th]} \BibitemShut
  {NoStop}%
\bibitem [{\citenamefont {Kim}\ and\ \citenamefont {Yi}(2024)}]{Kim:2024ewt}%
  \BibitemOpen
  \bibfield  {author} {\bibinfo {author} {\bibfnamefont {Taeyeon}\ \bibnamefont
  {Kim}}\ and\ \bibinfo {author} {\bibfnamefont {Piljin}\ \bibnamefont {Yi}},\
  }\bibfield  {title} {\enquote {\bibinfo {title} {{Lie, Noether, Kosmann, and
  Diffeomorphism Anomalies Redux}},}\ }\href@noop {} {\  (\bibinfo {year}
  {2024})},\ \Eprint {http://arxiv.org/abs/2412.03667} {arXiv:2412.03667
  [hep-th]} \BibitemShut {NoStop}%
\bibitem [{\citenamefont {Kosmann}(1966)}]{Kosmann:1966}%
  \BibitemOpen
  \bibfield  {author} {\bibinfo {author} {\bibfnamefont {Yvette}\ \bibnamefont
  {Kosmann}},\ }\bibfield  {title} {\enquote {\bibinfo {title}
  {D{\'e}riv{\'e}es de lie des spineurs. applications},}\ }\href@noop {}
  {\bibfield  {journal} {\bibinfo  {journal} {Comptes Rendus Hebdomadaires des
  Seances de l academie des Sciences Serie A}\ }\textbf {\bibinfo {volume}
  {262}},\ \bibinfo {pages} {394} (\bibinfo {year} {1966})}\BibitemShut
  {NoStop}%
\bibitem [{\citenamefont {Weyl}(1929)}]{Weyl:1929fm}%
  \BibitemOpen
  \bibfield  {author} {\bibinfo {author} {\bibfnamefont {H.}~\bibnamefont
  {Weyl}},\ }\bibfield  {title} {\enquote {\bibinfo {title} {{Electron and
  Gravitation. 1. (In German)}},}\ }\href {\doibase 10.1007/BF01339504}
  {\bibfield  {journal} {\bibinfo  {journal} {Z. Phys.}\ }\textbf {\bibinfo
  {volume} {56}},\ \bibinfo {pages} {330--352} (\bibinfo {year}
  {1929})}\BibitemShut {NoStop}%
\bibitem [{\citenamefont {Cartan}\ and\ \citenamefont
  {Mercier}(1981)}]{Cartan:1966}%
  \BibitemOpen
  \bibfield  {author} {\bibinfo {author} {\bibfnamefont {E.}~\bibnamefont
  {Cartan}}\ and\ \bibinfo {author} {\bibfnamefont {A.}~\bibnamefont
  {Mercier}},\ }\href {https://books.google.com/books?id=AEZ1h7Cg3cwC} {\emph
  {\bibinfo {title} {The Theory of Spinors}}},\ Dover Books on Mathematics\
  (\bibinfo  {publisher} {Dover Publications},\ \bibinfo {year}
  {1981})\BibitemShut {NoStop}%
\bibitem [{\citenamefont {Brading}\ and\ \citenamefont
  {Brown}(2000)}]{Brading:2000hc}%
  \BibitemOpen
  \bibfield  {author} {\bibinfo {author} {\bibfnamefont {Katherine}\
  \bibnamefont {Brading}}\ and\ \bibinfo {author} {\bibfnamefont {Harvey~R.}\
  \bibnamefont {Brown}},\ }\bibfield  {title} {\enquote {\bibinfo {title}
  {{Noether's theorems and gauge symmetries}},}\ }\href@noop {} {\  (\bibinfo
  {year} {2000})},\ \Eprint {http://arxiv.org/abs/hep-th/0009058}
  {arXiv:hep-th/0009058} \BibitemShut {NoStop}%
\bibitem [{\citenamefont {Brout}\ and\ \citenamefont
  {Englert}(1966)}]{Brout:1966oea}%
  \BibitemOpen
  \bibfield  {author} {\bibinfo {author} {\bibfnamefont {R.}~\bibnamefont
  {Brout}}\ and\ \bibinfo {author} {\bibfnamefont {F.}~\bibnamefont
  {Englert}},\ }\bibfield  {title} {\enquote {\bibinfo {title} {{Gravitational
  Ward Identity and the Principle of Equivalence}},}\ }\href {\doibase
  10.1103/PhysRev.141.1231} {\bibfield  {journal} {\bibinfo  {journal} {Phys.
  Rev.}\ }\textbf {\bibinfo {volume} {141}},\ \bibinfo {pages} {1231--1232}
  (\bibinfo {year} {1966})}\BibitemShut {NoStop}%
\bibitem [{\citenamefont {Freese}\ and\ \citenamefont
  {Clo\"et}(2019)}]{Freese:2019bhb}%
  \BibitemOpen
  \bibfield  {author} {\bibinfo {author} {\bibfnamefont {Adam}\ \bibnamefont
  {Freese}}\ and\ \bibinfo {author} {\bibfnamefont {Ian~C}\ \bibnamefont
  {Clo\"et}},\ }\bibfield  {title} {\enquote {\bibinfo {title} {{Gravitational
  form factors of light mesons}},}\ }\href {\doibase
  10.1103/PhysRevC.100.015201} {\bibfield  {journal} {\bibinfo  {journal}
  {Phys. Rev. C}\ }\textbf {\bibinfo {volume} {100}},\ \bibinfo {pages}
  {015201} (\bibinfo {year} {2019})},\ \bibinfo {note} {[Erratum: Phys.Rev.C
  105, 059901 (2022)]},\ \Eprint {http://arxiv.org/abs/1903.09222}
  {arXiv:1903.09222 [nucl-th]} \BibitemShut {NoStop}%
\bibitem [{\citenamefont {Gupta}(1950)}]{Gupta:1949rh}%
  \BibitemOpen
  \bibfield  {author} {\bibinfo {author} {\bibfnamefont {Suraj~N.}\
  \bibnamefont {Gupta}},\ }\bibfield  {title} {\enquote {\bibinfo {title}
  {{Theory of longitudinal photons in quantum electrodynamics}},}\ }\href
  {\doibase 10.1088/0370-1298/63/7/301} {\bibfield  {journal} {\bibinfo
  {journal} {Proc. Phys. Soc. A}\ }\textbf {\bibinfo {volume} {63}},\ \bibinfo
  {pages} {681--691} (\bibinfo {year} {1950})}\BibitemShut {NoStop}%
\bibitem [{\citenamefont {Bleuler}(1950)}]{Bleuler:1950cy}%
  \BibitemOpen
  \bibfield  {author} {\bibinfo {author} {\bibfnamefont {K.}~\bibnamefont
  {Bleuler}},\ }\bibfield  {title} {\enquote {\bibinfo {title} {{A New method
  of treatment of the longitudinal and scalar photons}},}\ }\href@noop {}
  {\bibfield  {journal} {\bibinfo  {journal} {Helv. Phys. Acta}\ }\textbf
  {\bibinfo {volume} {23}},\ \bibinfo {pages} {567--586} (\bibinfo {year}
  {1950})}\BibitemShut {NoStop}%
\bibitem [{\citenamefont {Proca}(1936)}]{Proca:1936fbw}%
  \BibitemOpen
  \bibfield  {author} {\bibinfo {author} {\bibfnamefont {Alexandru}\
  \bibnamefont {Proca}},\ }\bibfield  {title} {\enquote {\bibinfo {title} {{Sur
  la theorie ondulatoire des electrons positifs et negatifs}},}\ }\href
  {\doibase 10.1051/jphysrad:0193600708034700} {\bibfield  {journal} {\bibinfo
  {journal} {J. Phys. Radium}\ }\textbf {\bibinfo {volume} {7}},\ \bibinfo
  {pages} {347--353} (\bibinfo {year} {1936})}\BibitemShut {NoStop}%
\bibitem [{\citenamefont {Stueckelberg}(1938)}]{Stueckelberg:1938hvi}%
  \BibitemOpen
  \bibfield  {author} {\bibinfo {author} {\bibfnamefont {E.~C.~G.}\
  \bibnamefont {Stueckelberg}},\ }\bibfield  {title} {\enquote {\bibinfo
  {title} {{Interaction energy in electrodynamics and in the field theory of
  nuclear forces}},}\ }\href {\doibase 10.5169/seals-110852} {\bibfield
  {journal} {\bibinfo  {journal} {Helv. Phys. Acta}\ }\textbf {\bibinfo
  {volume} {11}},\ \bibinfo {pages} {225--244} (\bibinfo {year}
  {1938})}\BibitemShut {NoStop}%
\bibitem [{\citenamefont {Kugo}\ and\ \citenamefont
  {Ojima}(1979)}]{Kugo:1979gm}%
  \BibitemOpen
  \bibfield  {author} {\bibinfo {author} {\bibfnamefont {Taichiro}\
  \bibnamefont {Kugo}}\ and\ \bibinfo {author} {\bibfnamefont {Izumi}\
  \bibnamefont {Ojima}},\ }\bibfield  {title} {\enquote {\bibinfo {title}
  {{Local Covariant Operator Formalism of Nonabelian Gauge Theories and Quark
  Confinement Problem}},}\ }\href {\doibase 10.1143/PTPS.66.1} {\bibfield
  {journal} {\bibinfo  {journal} {Prog. Theor. Phys. Suppl.}\ }\textbf
  {\bibinfo {volume} {66}},\ \bibinfo {pages} {1--130} (\bibinfo {year}
  {1979})}\BibitemShut {NoStop}%
\bibitem [{\citenamefont {Faddeev}\ and\ \citenamefont
  {Popov}(1967)}]{Faddeev:1967fc}%
  \BibitemOpen
  \bibfield  {author} {\bibinfo {author} {\bibfnamefont {L.~D.}\ \bibnamefont
  {Faddeev}}\ and\ \bibinfo {author} {\bibfnamefont {V.~N.}\ \bibnamefont
  {Popov}},\ }\bibfield  {title} {\enquote {\bibinfo {title} {{Feynman Diagrams
  for the Yang-Mills Field}},}\ }\href {\doibase 10.1016/0370-2693(67)90067-6}
  {\bibfield  {journal} {\bibinfo  {journal} {Phys. Lett. B}\ }\textbf
  {\bibinfo {volume} {25}},\ \bibinfo {pages} {29--30} (\bibinfo {year}
  {1967})}\BibitemShut {NoStop}%
\bibitem [{\citenamefont {Becchi}\ \emph {et~al.}(1975)\citenamefont {Becchi},
  \citenamefont {Rouet},\ and\ \citenamefont {Stora}}]{Becchi:1974md}%
  \BibitemOpen
  \bibfield  {author} {\bibinfo {author} {\bibfnamefont {C.}~\bibnamefont
  {Becchi}}, \bibinfo {author} {\bibfnamefont {A.}~\bibnamefont {Rouet}}, \
  and\ \bibinfo {author} {\bibfnamefont {R.}~\bibnamefont {Stora}},\ }\bibfield
   {title} {\enquote {\bibinfo {title} {{Renormalization of the Abelian
  Higgs-Kibble Model}},}\ }\href {\doibase 10.1007/BF01614158} {\bibfield
  {journal} {\bibinfo  {journal} {Commun. Math. Phys.}\ }\textbf {\bibinfo
  {volume} {42}},\ \bibinfo {pages} {127--162} (\bibinfo {year}
  {1975})}\BibitemShut {NoStop}%
\bibitem [{\citenamefont {Becchi}\ \emph {et~al.}(1976)\citenamefont {Becchi},
  \citenamefont {Rouet},\ and\ \citenamefont {Stora}}]{Becchi:1975nq}%
  \BibitemOpen
  \bibfield  {author} {\bibinfo {author} {\bibfnamefont {C.}~\bibnamefont
  {Becchi}}, \bibinfo {author} {\bibfnamefont {A.}~\bibnamefont {Rouet}}, \
  and\ \bibinfo {author} {\bibfnamefont {R.}~\bibnamefont {Stora}},\ }\bibfield
   {title} {\enquote {\bibinfo {title} {{Renormalization of Gauge Theories}},}\
  }\href {\doibase 10.1016/0003-4916(76)90156-1} {\bibfield  {journal}
  {\bibinfo  {journal} {Annals Phys.}\ }\textbf {\bibinfo {volume} {98}},\
  \bibinfo {pages} {287--321} (\bibinfo {year} {1976})}\BibitemShut {NoStop}%
\bibitem [{\citenamefont {Tyutin}(1975)}]{Tyutin:1975qk}%
  \BibitemOpen
  \bibfield  {author} {\bibinfo {author} {\bibfnamefont {I.~V.}\ \bibnamefont
  {Tyutin}},\ }\bibfield  {title} {\enquote {\bibinfo {title} {{Gauge
  Invariance in Field Theory and Statistical Physics in Operator Formalism}},}\
  }\href@noop {} {\  (\bibinfo {year} {1975})},\ \Eprint
  {http://arxiv.org/abs/0812.0580} {arXiv:0812.0580 [hep-th]} \BibitemShut
  {NoStop}%
\bibitem [{\citenamefont {Collins}(1986)}]{Collins:1984xc}%
  \BibitemOpen
  \bibfield  {author} {\bibinfo {author} {\bibfnamefont {John~C.}\ \bibnamefont
  {Collins}},\ }\href {\doibase 10.1017/CBO9780511622656} {\emph {\bibinfo
  {title} {{Renormalization}}}},\ \bibinfo {series} {Cambridge Monographs on
  Mathematical Physics}, Vol.~\bibinfo {volume} {26}\ (\bibinfo  {publisher}
  {Cambridge University Press},\ \bibinfo {address} {Cambridge},\ \bibinfo
  {year} {1986})\BibitemShut {NoStop}%
\bibitem [{\citenamefont {Iosifidis}\ \emph {et~al.}(2025)\citenamefont
  {Iosifidis}, \citenamefont {Karydas}, \citenamefont {Petkou},\ and\
  \citenamefont {Siampos}}]{Iosifidis:2025sjx}%
  \BibitemOpen
  \bibfield  {author} {\bibinfo {author} {\bibfnamefont {Damianos}\
  \bibnamefont {Iosifidis}}, \bibinfo {author} {\bibfnamefont {Manthos}\
  \bibnamefont {Karydas}}, \bibinfo {author} {\bibfnamefont {Anastasios}\
  \bibnamefont {Petkou}}, \ and\ \bibinfo {author} {\bibfnamefont
  {Konstantinos}\ \bibnamefont {Siampos}},\ }\bibfield  {title} {\enquote
  {\bibinfo {title} {{On the geometric origin of the energy-momentum tensor
  improvement terms}},}\ }\href@noop {} {\  (\bibinfo {year} {2025})},\ \Eprint
  {http://arxiv.org/abs/2503.21609} {arXiv:2503.21609 [hep-th]} \BibitemShut
  {NoStop}%
\bibitem [{\citenamefont {Hehl}\ \emph {et~al.}(1976)\citenamefont {Hehl},
  \citenamefont {Von Der~Heyde}, \citenamefont {Kerlick},\ and\ \citenamefont
  {Nester}}]{Hehl:1976kj}%
  \BibitemOpen
  \bibfield  {author} {\bibinfo {author} {\bibfnamefont {F.~W.}\ \bibnamefont
  {Hehl}}, \bibinfo {author} {\bibfnamefont {P.}~\bibnamefont {Von Der~Heyde}},
  \bibinfo {author} {\bibfnamefont {G.~D.}\ \bibnamefont {Kerlick}}, \ and\
  \bibinfo {author} {\bibfnamefont {J.~M.}\ \bibnamefont {Nester}},\ }\bibfield
   {title} {\enquote {\bibinfo {title} {{General Relativity with Spin and
  Torsion: Foundations and Prospects}},}\ }\href {\doibase
  10.1103/RevModPhys.48.393} {\bibfield  {journal} {\bibinfo  {journal} {Rev.
  Mod. Phys.}\ }\textbf {\bibinfo {volume} {48}},\ \bibinfo {pages} {393--416}
  (\bibinfo {year} {1976})}\BibitemShut {NoStop}%
\bibitem [{\citenamefont {Hammond}(2002)}]{Hammond:2002rm}%
  \BibitemOpen
  \bibfield  {author} {\bibinfo {author} {\bibfnamefont {R.~T.}\ \bibnamefont
  {Hammond}},\ }\bibfield  {title} {\enquote {\bibinfo {title} {{Torsion
  gravity}},}\ }\href {\doibase 10.1088/0034-4885/65/5/201} {\bibfield
  {journal} {\bibinfo  {journal} {Rept. Prog. Phys.}\ }\textbf {\bibinfo
  {volume} {65}},\ \bibinfo {pages} {599--649} (\bibinfo {year}
  {2002})}\BibitemShut {NoStop}%
\bibitem [{\citenamefont {Shapiro}(2002)}]{Shapiro:2001rz}%
  \BibitemOpen
  \bibfield  {author} {\bibinfo {author} {\bibfnamefont {I.~L.}\ \bibnamefont
  {Shapiro}},\ }\bibfield  {title} {\enquote {\bibinfo {title} {{Physical
  aspects of the space-time torsion}},}\ }\href {\doibase
  10.1016/S0370-1573(01)00030-8} {\bibfield  {journal} {\bibinfo  {journal}
  {Phys. Rept.}\ }\textbf {\bibinfo {volume} {357}},\ \bibinfo {pages} {113}
  (\bibinfo {year} {2002})},\ \Eprint {http://arxiv.org/abs/hep-th/0103093}
  {arXiv:hep-th/0103093} \BibitemShut {NoStop}%
\bibitem [{\citenamefont {Cartan}(1922)}]{cartan1922generalisation}%
  \BibitemOpen
  \bibfield  {author} {\bibinfo {author} {\bibfnamefont {{\'E}lie}\
  \bibnamefont {Cartan}},\ }\bibfield  {title} {\enquote {\bibinfo {title} {Sur
  une g{\'e}n{\'e}ralisation de la notion de courbure de riemann et les espaces
  {\`a} torsion},}\ }\href@noop {} {\bibfield  {journal} {\bibinfo  {journal}
  {Comptes Rendus, Ac. Sc. Paris}\ }\textbf {\bibinfo {volume} {174}},\
  \bibinfo {pages} {593--595} (\bibinfo {year} {1922})}\BibitemShut {NoStop}%
\bibitem [{\citenamefont {Kibble}(1961)}]{Kibble:1961ba}%
  \BibitemOpen
  \bibfield  {author} {\bibinfo {author} {\bibfnamefont {T.~W.~B.}\
  \bibnamefont {Kibble}},\ }\bibfield  {title} {\enquote {\bibinfo {title}
  {{Lorentz invariance and the gravitational field}},}\ }\href {\doibase
  10.1063/1.1703702} {\bibfield  {journal} {\bibinfo  {journal} {J. Math.
  Phys.}\ }\textbf {\bibinfo {volume} {2}},\ \bibinfo {pages} {212--221}
  (\bibinfo {year} {1961})}\BibitemShut {NoStop}%
\bibitem [{\citenamefont {Sciama}(1964)}]{Sciama:1964jqa}%
  \BibitemOpen
  \bibfield  {author} {\bibinfo {author} {\bibfnamefont {D.~W.}\ \bibnamefont
  {Sciama}},\ }\bibfield  {title} {\enquote {\bibinfo {title} {{The Physical
  Structure of General Relativity}},}\ }\href {\doibase
  10.1103/RevModPhys.36.463} {\bibfield  {journal} {\bibinfo  {journal} {Rev.
  Mod. Phys.}\ }\textbf {\bibinfo {volume} {36}},\ \bibinfo {pages} {463}
  (\bibinfo {year} {1964})}\BibitemShut {NoStop}%
\bibitem [{\citenamefont {Jockel}\ and\ \citenamefont
  {Menger}(2024)}]{Jockel:2024fps}%
  \BibitemOpen
  \bibfield  {author} {\bibinfo {author} {\bibfnamefont {C{\'e}dric}\
  \bibnamefont {Jockel}}\ and\ \bibinfo {author} {\bibfnamefont {Leon}\
  \bibnamefont {Menger}},\ }\bibfield  {title} {\enquote {\bibinfo {title}
  {{Effect of torsion on neutron star structure in Einstein-Cartan gravity}},}\
  }\href {\doibase 10.1103/PhysRevD.110.104022} {\bibfield  {journal} {\bibinfo
   {journal} {Phys. Rev. D}\ }\textbf {\bibinfo {volume} {110}},\ \bibinfo
  {pages} {104022} (\bibinfo {year} {2024})},\ \Eprint
  {http://arxiv.org/abs/2406.05851} {arXiv:2406.05851 [gr-qc]} \BibitemShut
  {NoStop}%
\bibitem [{\citenamefont {Andersson}\ and\ \citenamefont
  {Kokkotas}(1998)}]{Andersson:1997rn}%
  \BibitemOpen
  \bibfield  {author} {\bibinfo {author} {\bibfnamefont {Nils}\ \bibnamefont
  {Andersson}}\ and\ \bibinfo {author} {\bibfnamefont {Kostas~D.}\ \bibnamefont
  {Kokkotas}},\ }\bibfield  {title} {\enquote {\bibinfo {title} {{Towards
  gravitational wave asteroseismology}},}\ }\href {\doibase
  10.1046/j.1365-8711.1998.01840.x} {\bibfield  {journal} {\bibinfo  {journal}
  {Mon. Not. Roy. Astron. Soc.}\ }\textbf {\bibinfo {volume} {299}},\ \bibinfo
  {pages} {1059--1068} (\bibinfo {year} {1998})},\ \Eprint
  {http://arxiv.org/abs/gr-qc/9711088} {arXiv:gr-qc/9711088} \BibitemShut
  {NoStop}%
\bibitem [{\citenamefont {Benhar}\ \emph {et~al.}(2004)\citenamefont {Benhar},
  \citenamefont {Ferrari},\ and\ \citenamefont {Gualtieri}}]{Benhar:2004xg}%
  \BibitemOpen
  \bibfield  {author} {\bibinfo {author} {\bibfnamefont {Omar}\ \bibnamefont
  {Benhar}}, \bibinfo {author} {\bibfnamefont {Valeria}\ \bibnamefont
  {Ferrari}}, \ and\ \bibinfo {author} {\bibfnamefont {Leonardo}\ \bibnamefont
  {Gualtieri}},\ }\bibfield  {title} {\enquote {\bibinfo {title}
  {{Gravitational wave asteroseismology revisited}},}\ }\href {\doibase
  10.1103/PhysRevD.70.124015} {\bibfield  {journal} {\bibinfo  {journal} {Phys.
  Rev. D}\ }\textbf {\bibinfo {volume} {70}},\ \bibinfo {pages} {124015}
  (\bibinfo {year} {2004})},\ \Eprint {http://arxiv.org/abs/astro-ph/0407529}
  {arXiv:astro-ph/0407529} \BibitemShut {NoStop}%
\bibitem [{\citenamefont {Tsui}\ and\ \citenamefont
  {Leung}(2005)}]{Tsui:2005zf}%
  \BibitemOpen
  \bibfield  {author} {\bibinfo {author} {\bibfnamefont {L.~K.}\ \bibnamefont
  {Tsui}}\ and\ \bibinfo {author} {\bibfnamefont {Pui-Tang}\ \bibnamefont
  {Leung}},\ }\bibfield  {title} {\enquote {\bibinfo {title} {{Probing the
  interior of neutron stars with gravitational waves}},}\ }\href {\doibase
  10.1103/PhysRevLett.95.151101} {\bibfield  {journal} {\bibinfo  {journal}
  {Phys. Rev. Lett.}\ }\textbf {\bibinfo {volume} {95}},\ \bibinfo {pages}
  {151101} (\bibinfo {year} {2005})},\ \Eprint
  {http://arxiv.org/abs/astro-ph/0506681} {arXiv:astro-ph/0506681} \BibitemShut
  {NoStop}%
\bibitem [{\citenamefont {Sotani}\ \emph {et~al.}(2012)\citenamefont {Sotani},
  \citenamefont {Nakazato}, \citenamefont {Iida},\ and\ \citenamefont
  {Oyamatsu}}]{Sotani:2012qc}%
  \BibitemOpen
  \bibfield  {author} {\bibinfo {author} {\bibfnamefont {Hajime}\ \bibnamefont
  {Sotani}}, \bibinfo {author} {\bibfnamefont {Ken'ichiro}\ \bibnamefont
  {Nakazato}}, \bibinfo {author} {\bibfnamefont {Kei}\ \bibnamefont {Iida}}, \
  and\ \bibinfo {author} {\bibfnamefont {Kazuhiro}\ \bibnamefont {Oyamatsu}},\
  }\bibfield  {title} {\enquote {\bibinfo {title} {{Probing the Equation of
  State of Nuclear Matter via Neutron Star Asteroseismology}},}\ }\href
  {\doibase 10.1103/PhysRevLett.108.201101} {\bibfield  {journal} {\bibinfo
  {journal} {Phys. Rev. Lett.}\ }\textbf {\bibinfo {volume} {108}},\ \bibinfo
  {pages} {201101} (\bibinfo {year} {2012})},\ \Eprint
  {http://arxiv.org/abs/1202.6242} {arXiv:1202.6242 [astro-ph.HE]} \BibitemShut
  {NoStop}%
\bibitem [{\citenamefont {Weinberg}(2005)}]{Weinberg:1995mt}%
  \BibitemOpen
  \bibfield  {author} {\bibinfo {author} {\bibfnamefont {Steven}\ \bibnamefont
  {Weinberg}},\ }\href {\doibase 10.1017/CBO9781139644167} {\emph {\bibinfo
  {title} {{The Quantum theory of fields. Vol. 1: Foundations}}}}\ (\bibinfo
  {publisher} {Cambridge University Press},\ \bibinfo {year}
  {2005})\BibitemShut {NoStop}%
\bibitem [{\citenamefont {Hehl}\ and\ \citenamefont
  {Obukhov}(2003)}]{hehl2003foundations}%
  \BibitemOpen
  \bibfield  {author} {\bibinfo {author} {\bibfnamefont {F.W.}\ \bibnamefont
  {Hehl}}\ and\ \bibinfo {author} {\bibfnamefont {Y.N.}\ \bibnamefont
  {Obukhov}},\ }\href {https://books.google.com/books?id=48-hHXL-CYUC} {\emph
  {\bibinfo {title} {Foundations of Classical Electrodynamics: Charge, Flux,
  and Metric}}},\ Progress in Mathematical Physics\ (\bibinfo  {publisher}
  {Birkh{\"a}user Boston},\ \bibinfo {year} {2003})\BibitemShut {NoStop}%
\bibitem [{\citenamefont {Lorce}(2013)}]{Lorce:2012rr}%
  \BibitemOpen
  \bibfield  {author} {\bibinfo {author} {\bibfnamefont {Cedric}\ \bibnamefont
  {Lorce}},\ }\bibfield  {title} {\enquote {\bibinfo {title} {{Geometrical
  approach to the proton spin decomposition}},}\ }\href {\doibase
  10.1103/PhysRevD.87.034031} {\bibfield  {journal} {\bibinfo  {journal} {Phys.
  Rev. D}\ }\textbf {\bibinfo {volume} {87}},\ \bibinfo {pages} {034031}
  (\bibinfo {year} {2013})},\ \Eprint {http://arxiv.org/abs/1205.6483}
  {arXiv:1205.6483 [hep-ph]} \BibitemShut {NoStop}%
\bibitem [{\citenamefont {Hehl}(2014)}]{Hehl:2014eja}%
  \BibitemOpen
  \bibfield  {author} {\bibinfo {author} {\bibfnamefont {Friedrich~W.}\
  \bibnamefont {Hehl}},\ }\bibfield  {title} {\enquote {\bibinfo {title} {{On
  energy-momentum and spin/helicity of quark and gluon fields}},}\ }in\
  \href@noop {} {\emph {\bibinfo {booktitle} {{15th Workshop on High Energy
  Spin Physics}}}}\ (\bibinfo {year} {2014})\ \Eprint
  {http://arxiv.org/abs/1402.0261} {arXiv:1402.0261 [gr-qc]} \BibitemShut
  {NoStop}%
\bibitem [{\citenamefont {Bohm}(1989)}]{bohm1989quantum}%
  \BibitemOpen
  \bibfield  {author} {\bibinfo {author} {\bibfnamefont {D.}~\bibnamefont
  {Bohm}},\ }\href {https://books.google.com/books?id=GKVGDwAAQBAJ} {\emph
  {\bibinfo {title} {Quantum Theory}}},\ Dover books in science and
  mathematics\ (\bibinfo  {publisher} {Dover Publications},\ \bibinfo {year}
  {1989})\BibitemShut {NoStop}%
\bibitem [{\citenamefont {Cohen-Tannoudji}\ \emph {et~al.}(2019)\citenamefont
  {Cohen-Tannoudji}, \citenamefont {Diu},\ and\ \citenamefont
  {Lalo{\"e}}}]{cohen2019quantum}%
  \BibitemOpen
  \bibfield  {author} {\bibinfo {author} {\bibfnamefont {C.}~\bibnamefont
  {Cohen-Tannoudji}}, \bibinfo {author} {\bibfnamefont {B.}~\bibnamefont
  {Diu}}, \ and\ \bibinfo {author} {\bibfnamefont {F.}~\bibnamefont
  {Lalo{\"e}}},\ }\href {https://books.google.com/books?id=tVI_EAAAQBAJ} {\emph
  {\bibinfo {title} {Quantum Mechanics, Volume 1: Basic Concepts, Tools, and
  Applications}}}\ (\bibinfo  {publisher} {Wiley},\ \bibinfo {year}
  {2019})\BibitemShut {NoStop}%
\bibitem [{\citenamefont {Sakurai}\ and\ \citenamefont
  {Napolitano}(2020)}]{Sakurai:2011zz}%
  \BibitemOpen
  \bibfield  {author} {\bibinfo {author} {\bibfnamefont {Jun~John}\
  \bibnamefont {Sakurai}}\ and\ \bibinfo {author} {\bibfnamefont {Jim}\
  \bibnamefont {Napolitano}},\ }\href {\doibase 10.1017/9781108587280} {\emph
  {\bibinfo {title} {{Modern Quantum Mechanics}}}},\ Quantum physics, quantum
  information and quantum computation\ (\bibinfo  {publisher} {Cambridge
  University Press},\ \bibinfo {year} {2020})\BibitemShut {NoStop}%
\bibitem [{\citenamefont {Bates}\ and\ \citenamefont
  {Weinstein}(1997)}]{bates1997lectures}%
  \BibitemOpen
  \bibfield  {author} {\bibinfo {author} {\bibfnamefont {S.}~\bibnamefont
  {Bates}}\ and\ \bibinfo {author} {\bibfnamefont {A.}~\bibnamefont
  {Weinstein}},\ }\href
  {https://ncatlab.org/nlab/files/BatesWeinstein-Lectures.pdf} {\emph {\bibinfo
  {title} {Lectures on the Geometry of Quantization}}},\ Berkeley mathematics
  lecture notes\ (\bibinfo  {publisher} {American Mathematical Society},\
  \bibinfo {year} {1997})\BibitemShut {NoStop}%
\bibitem [{\citenamefont {Bongers}(2014)}]{bongers2014geometric}%
  \BibitemOpen
  \bibfield  {author} {\bibinfo {author} {\bibfnamefont {S.R.}\ \bibnamefont
  {Bongers}},\ }\emph {\bibinfo {title} {Geometric quantization of symplectic
  and Poisson manifolds}},\ \href
  {https://studenttheses.uu.nl/handle/20.500.12932/16148} {Master's thesis}
  (\bibinfo {year} {2014})\BibitemShut {NoStop}%
\bibitem [{\citenamefont {Puk{\'a}nszky}(1964)}]{pukanszky1964plancherel}%
  \BibitemOpen
  \bibfield  {author} {\bibinfo {author} {\bibfnamefont {Lajos}\ \bibnamefont
  {Puk{\'a}nszky}},\ }\bibfield  {title} {\enquote {\bibinfo {title} {The
  plancherel formula for the universal covering group of sl(r,2)},}\
  }\href@noop {} {\bibfield  {journal} {\bibinfo  {journal} {Mathematische
  Annalen}\ }\textbf {\bibinfo {volume} {156}},\ \bibinfo {pages} {96--143}
  (\bibinfo {year} {1964})}\BibitemShut {NoStop}%
\bibitem [{\citenamefont {Kitaev}(2017)}]{Kitaev:2017hnr}%
  \BibitemOpen
  \bibfield  {author} {\bibinfo {author} {\bibfnamefont {Alexei}\ \bibnamefont
  {Kitaev}},\ }\bibfield  {title} {\enquote {\bibinfo {title} {{Notes on
  $\widetilde{\mathrm{SL}}(2,\mathbb{R})$ representations}},}\ }\href@noop {}
  {\  (\bibinfo {year} {2017})},\ \Eprint {http://arxiv.org/abs/1711.08169}
  {arXiv:1711.08169 [hep-th]} \BibitemShut {NoStop}%
\bibitem [{\citenamefont {Weissman}(2023)}]{Weismann:2023}%
  \BibitemOpen
  \bibfield  {author} {\bibinfo {author} {\bibfnamefont {Martin}\ \bibnamefont
  {Weissman}},\ }\bibfield  {title} {\enquote {\bibinfo {title} {What is... a
  metaplectic group?}}\ }\href {\doibase 10.1090/noti2687} {\bibfield
  {journal} {\bibinfo  {journal} {Notices of the American Mathematical
  Society}\ }\textbf {\bibinfo {volume} {70}},\ \bibinfo {pages} {1} (\bibinfo
  {year} {2023})}\BibitemShut {NoStop}%
\bibitem [{\citenamefont {Weil}\ \emph {et~al.}(1964)\citenamefont {Weil} \emph
  {et~al.}}]{weil1964certains}%
  \BibitemOpen
  \bibfield  {author} {\bibinfo {author} {\bibfnamefont {Andr{\'e}}\
  \bibnamefont {Weil}} \emph {et~al.},\ }\bibfield  {title} {\enquote {\bibinfo
  {title} {Sur certains groupes d’op{\'e}rateurs unitaires},}\ }\href@noop {}
  {\bibfield  {journal} {\bibinfo  {journal} {Acta math}\ }\textbf {\bibinfo
  {volume} {111}},\ \bibinfo {pages} {14} (\bibinfo {year} {1964})}\BibitemShut
  {NoStop}%
\bibitem [{\citenamefont {Ogievetsky}\ and\ \citenamefont
  {Polubarinov}(1965)}]{Ogievetsky:1965ii}%
  \BibitemOpen
  \bibfield  {author} {\bibinfo {author} {\bibfnamefont {V.~I.}\ \bibnamefont
  {Ogievetsky}}\ and\ \bibinfo {author} {\bibfnamefont {I.~V.}\ \bibnamefont
  {Polubarinov}},\ }\bibfield  {title} {\enquote {\bibinfo {title} {{Spinors in
  gravitation theory}},}\ }\href@noop {} {\bibfield  {journal} {\bibinfo
  {journal} {Sov. Phys. JETP}\ }\textbf {\bibinfo {volume} {21}},\ \bibinfo
  {pages} {1093--1100} (\bibinfo {year} {1965})}\BibitemShut {NoStop}%
\bibitem [{\citenamefont {Pitts}(2012)}]{Pitts:2011jv}%
  \BibitemOpen
  \bibfield  {author} {\bibinfo {author} {\bibfnamefont {J.~Brian}\
  \bibnamefont {Pitts}},\ }\bibfield  {title} {\enquote {\bibinfo {title} {{The
  Nontriviality of Trivial General Covariance: How Electrons Restrict 'Time'
  Coordinates, Spinors (Almost) Fit into Tensor Calculus, and 7/16 of a Tetrad
  Is Surplus Structure}},}\ }\href {\doibase 10.1016/j.shpsb.2011.11.001}
  {\bibfield  {journal} {\bibinfo  {journal} {Stud. Hist. Phil. Sci. B}\
  }\textbf {\bibinfo {volume} {43}},\ \bibinfo {pages} {1--24} (\bibinfo {year}
  {2012})},\ \Eprint {http://arxiv.org/abs/1111.4586} {arXiv:1111.4586 [gr-qc]}
  \BibitemShut {NoStop}%
\end{thebibliography}%

\end{document}